\documentstyle[12pt,aaspp4]{article}  
\begin{document}
\title {ENERGETICS OF TeV BLAZARS AND PHYSICAL CONSTRAINTS
ON THEIR EMISSION REGIONS}
\author { MOTOKI KINO \altaffilmark{1,2}, FUMIO
TAKAHARA\altaffilmark{1}
 and MASAAKI KUSUNOSE \altaffilmark{3}}
\affil {\altaffilmark{1}Department of Earth and Space Science,
Graduate School of Science,
Osaka University, \\ Toyonaka, Osaka 560-0043, Japan;\\
kino@vega.ess.sci.osaka-u.ac.jp, takahara@vega.ess.sci.osaka-u.ac.jp}
\affil {\altaffilmark{2}
Astronomical Institute, Graduate School of Science,
Tohoku University, \\ Aoba-ku, Sendai 980-8578, Japan;
kino@astr.tohoku.ac.jp}
\affil {\altaffilmark{3}
Department of Physics, School of Science,
Kwansei Gakuin University, \\ Nishinomiya 662-8501, Japan;
kusunose@kwansei.ac.jp}

\begin{abstract}
Using multi-frequency spectra from TeV blazars in quiescent states,
we obtain the physical parameters of the emission region of blazars
within the framework of the one-zone synchrotron self-Compton (SSC) model.
We numerically calculate the steady-state energy spectra of electrons
by self-consistently taking into account the effects of radiative cooling
with a proper account of the Klein-Nishina effects.
Here electrons are assumed to be injected with a power-law spectrum
and to escape on a finite time scale, which naturally leads to the
existence of a break energy scale.
Although we do not use time variabilities but utilize
a model of electron escape to constrain the size of the
emission region, the resultant size turns out to be similar
to that obtained based on time variabilities.
Through detailed comparison of the predicted emission spectra
with observations,
we find that for Mrk 421, Mrk 501, and PKS 2155--304,
the energy density of relativistic electrons
is about an order of magnitude larger than that of magnetic fields
with an uncertainty within a factor of a few.

\end{abstract}


\section{INTRODUCTION}
Blazars comprising of BL Lac objects and optically violent variable
quasars are characterized by rapid time variation of the energy flux,
large and variable polarization, and featureless continuum spectra
(e.g., Urry \& Padovani 1995).
These characteristics are considered to be the result of beamed emission
from
relativistic jets seen end-on
(e.g., Blandford \& K\"{o}nigl 1979).
The discovery of strong $\gamma$-ray emission from blazars in the
GeV band by EGRET on {\it Compton Gamma
Ray Observatory} is one of the most important issues for active galactic
nuclei (AGNs), because more than 60 AGNs detected by EGRET are all
blazar type and no identifications as other types of AGNs such as
Seyfert galaxies have been reported 
(Mukherjee et al. 1997)
except for a probable detection from the radio galaxy 
Centaurus A 
(Hartman et al. 1999).
Multi-frequency observations
have revealed that
broad band continuum spectra of blazars consist of two components;
the low energy component from radio to optical/UV band sometimes extending
to X-ray band is by synchrotron radiation,
while the high energy component
from X-ray to $\gamma$-ray is due to the inverse Compton scattering of soft
photons
(e.g., Kubo et al. 1998):
various soft photon sources have been proposed ranging
from synchrotron photons to photons from accretion disks, either
direct or reprocessed
(e.g., Sikora, Begelman \& Rees 1994; Inoue \& Takahara 1996;
Dermer \& Schlickeiser 1993; Blandford \& Levinson 1995; Ghisellini \&
Madau 1996).
One of the most important aspects of multi-frequency observations of
blazars is to probe the energetics of relativistic jets.  From this point,
a few blazars from which TeV $\gamma$-rays have been
detected (Mrk 421, Mrk 501, PKS 2155--304, and 1ES 2344+514) are especially
important, because these TeV blazars are relatively
less luminous and pure synchrotron self-Compton model can be best applied.

Up to now, source parameters of TeV blazars have been estimated in a
variety of ways
(e.g., Bednarek \& Protheroe 1997; Tavecchio et al. 1998; Kataoka et
al. 2000).
With regard to the basic energetics, however,
surprisingly little attention has been paid to the energetics of
electrons, which should be the central
concern for the theoretical understanding
of the production and bulk acceleration of
relativistic jets.  Based on an analytic estimate from TeV blazar
observations, one of the present authors argued that relativistic
electrons dominate over magnetic fields in energy densities in
relativistic jets of blazars 
(Takahara 1997).
The purpose of this study is to estimate the energy
densities more quantitatively using a numerical code which self-consistently
solves for electron spectra suffering from injection, escape, and radiative
cooling and photon spectra with proper account of the Klein-Nishina effects
(e.g., Mastichiadis \& Kirk 1997; Li \& Kusunose 2000).
Although recent observations show that
even in low activity states 
$\gamma$-ray spectra extend above 1 TeV 
(e.g., Aharonian et al. 2001), 
in this paper, we restrict our attention to the quiescent
states and treat the $\gamma$-rays below 1TeV 
in the first step neglecting the correction for
the absorption of TeV $\gamma$-rays due to
Cosmic Infra-Red Background (CIB).
We shortly discuss  CIB absorption effects in \S \ref{dis}.
In a future research,
we will separately examine this issue 
including the flaring states 
where $\gamma$-ray spectrum
clearly extends up to 10 TeV. 

The rest of the paper is organized as follows.
In \S \ref{model}, we introduce the synchrotron self-Compton (SSC) model,
and we describe our numerical treatment for solving the kinetic equations of
photons and electrons.  We also show the relation between model parameters
and typical observables to help the search for the parameter set of
the best fit model in numerical calculations.
In \S \ref{energy}, we discuss the analytic estimation of
the ratio of the energy density of
relativistic electrons to that of magnetic fields.
In \S \ref{result}, we show the numerical results of spectral fitting
applied to three TeV blazars, i.e., Mrk 421, Mrk 501, and PKS2155--304.
(1ES2344+514 is omitted because less data are available at present.)
Finally in \S \ref{dis}, we summarize our main results and discuss
some related issues.

\section{ONE-ZONE SSC MODEL}\label{model}

\subsection{Basic Assumptions}

Non-thermal emission from TeV blazars is divided into two components, i.e.,
low energy synchrotron component extending from radio to X-rays,
and high energy inverse Compton component extending from hard X-rays
to TeV $\gamma$-rays.
Here the seed photons of inverse Compton scattering are the synchrotron
photons in the same emission region.
This SSC model has been very successful in describing the observed
multi-frequency spectra 
(e.g., Jones et al. 1974; Maraschi, Ghisellini, \& Celotti 1992).

Further assumptions used in the present work are that
(1) emission region is one zone with a characteristic size $R$
and is moving at a relativistic speed $\beta$ in units of the light speed
and that
(2) both relativistic electrons and photons are
isotropic in the source frame.
Beaming (Doppler) factor is given by $\delta=1/[\Gamma(1-\beta\cos\theta)]$,
where $\theta$ is the angle between the line of sight and the direction
of the relativistic jet and $\Gamma$ is the bulk Lorentz factor of
the emission region in the jet.
When the observer lies within the angle of $\theta\sim1/\Gamma$,
we obtain $\Gamma\sim\delta$. 
The Hubble constant is assumed to be $75\ \rm km\ s^{-1}\ Mpc^{-1}$.
 Throughout this paper we use these
approximations.

\subsection{Numerical Approach}

Most of previous calculations are either semi-analytic or done without
the inverse Compton process self-consistently.
In our numerical code, to obtain the consistent spectra of photons
and relativistic electrons, we calculate the kinetic equations
of electrons and photons
self-consistently including the exact inverse Compton process
within the continuous energy loss approximation.

The kinetic equation describing the time evolution of
the electron distribution is given by
\begin{eqnarray}\label{ekin}
 \frac{\partial n_{\rm e}(\gamma,t)}{\partial t}
   +\frac{ n_{\rm e}(\gamma,t)}{t_{\rm e,esc}}
=   -\frac{\partial }{\partial \gamma}
    \left[({\dot \gamma}_{\rm syn}
                +{\dot \gamma}_{\rm ssc})n_{\rm e}(\gamma,t)\right]
     +{\cal Q}_{\rm e,inj}(\gamma,t),
\end{eqnarray}
where $\gamma$ is the electron Lorentz factor and
$n_{\rm e}$ is the electron number density per $\gamma$;
${\dot \gamma}_{\rm syn}$ and ${\dot \gamma}_{\rm ssc}$ are
the cooling rates of 
synchrotron and inverse Compton emission,
respectively;
$t_{\rm e,esc}$ is the effective escape time of the electrons,
which is identified as the time scale of 
the adiabatic expansion loss 
(Mastichiadis \& Kirk 1997).
For simplicity, in all the following numerical calculations, we set
$t_{\rm e,esc}=3t_{\rm dyn}$, where $t_{\rm dyn}\equiv R/c$.
Next we adopt an injection spectrum
\begin{equation}
{\cal Q}_{\rm e,inj}=q_{\rm e}\gamma^{-s}e^{-\gamma/\gamma_{\rm max}}
\quad {\rm for} \ \gamma_{\rm min} < \gamma,
\end{equation}
where $\gamma_{\rm max}$ and $\gamma_{\rm min}$ are, respectively,
the maximum and minimum Lorentz factors of the electrons,
$q_{\rm e}$ is the normalization factor, and $s$ is the power-law index.
As for the injection mechanism, we implicitly assume the first order
Fermi acceleration
(e.g., Blandford \& Eichler 1987).
The synchrotron emissivity and absorption
coefficient are calculated based on
\cite{rm84}
and \cite{cs86}.
In the calculation of the inverse Compton scattering,
we use the exact Klein-Nishina cross section and scattering probability
of \cite{j68}
and \cite{cb90}.

The kinetic equation of photons is given by
\begin{eqnarray} \label{pkin}
  \frac{\partial n_{\rm ph}(\epsilon,t)}{\partial t}
    +\frac{ n_{\rm ph}(\epsilon,t)}{t_{\rm ph,esc}} =
     {\dot n}_{\rm IC}(\epsilon,t)
    + {\dot n}_{\rm syn}(\epsilon,t)
\end{eqnarray}
where $\epsilon$ is the dimensionless photon energy normalized by
$m_{\rm e}c^{2}$ with $m_{\rm e}$ being the electron mass,
$n_{\rm ph}$ is the photon number density
per unit energy $\epsilon$,
$t_{\rm ph,esc}$ is the escape time of photons from the emission region,
which is taken as $t_{\rm ph,esc}=R/c$ in the optically thin limit.
${\dot n}_{\rm IC}(\epsilon,t)$ and
${\dot n}_{\rm syn}(\epsilon,t)$ are the production rate of 
inverse Compton and synchrotron photons per unit energy
$\epsilon$, respectively.

In order to obtain quiescent state spectra, calculations are done
up to $15t_{\rm dyn}$, which is long enough to reach a steady state.
The physical quantities in the source frame can be converted to those in the
observer frame using the relations such as
$\epsilon_{\rm o}=\epsilon_{\rm s}\delta/(1+z)$
and $dt_{\rm o}=dt_{\rm s} (1+z)/\delta$,
where subscripts ${\rm o}$ and ${\rm s}$ express the quantity in the observer
and
source frame, respectively, and $z$ is the redshift of the source.

In this model, there are seven parameters to be determined
by the comparison of predicted and observed photon spectra.
They are:
$R$ the size of the emission region,
$B$ the magnetic field strength,
$\delta$ the beaming factor,
$\gamma_{\rm max}$ the maximum Lorentz factor,
$\gamma_{\rm min}$ the minimum Lorentz factor,
$q_{\rm e}$ the injection rate of electrons,
and
$s$ the power-law index of the injected electron spectrum.
Among them, $\gamma_{\rm min}$ is not easily constrained by
spectral fitting and it is taken to be 10 in all the numerical calculations.
Although $\gamma_{\rm min}$ hardly affects the radiation spectra, it is
important for probing the energy and number densities of
relativistic electrons and thus the matter content of the relativistic jets,
and it has been a matter of debate
(e.g., Reynolds et al. 1996; Wardle et al. 1998; Hirotani et al. 1999).
We discuss the effect of changing the value of $\gamma_{\rm min}$
in \S \ref{dis}.

\subsection{Analytic Estimate}

Before we present numerical results, we describe some analytic estimates
which provide useful insight into the physics behind the
relationship between the model parameters and typical observables
and we later examine quantitatively to what extent simple analytic methods
are
accurate.
Using these relations, we constitute an analytic
estimate of model parameters which are then used
as a starting set of parameters for numerical calculations.

\subsubsection{Relation between the Model Parameters and Observables}

Among seven parameters, $\gamma_{\rm min}$ is least constrained and
taken to be 10 as was mentioned above.
The index $s$ is determined by the spectral shape
of the synchrotron radiation at low energies; specifically, the energy index
$\alpha$ between the radio and IR band is used to determine $s$ by
\begin{eqnarray}
s=2\alpha+1.
\end{eqnarray}
Other five parameters,
$R$, $B$, $\delta$, $\gamma_{\rm max}$, and $q_{\rm e}$ remain to be
determined.  Basically, the luminosities and typical frequencies of
synchrotron and inverse-Compton components give four constraints.
Remaining one can be taken to be the break frequency of the synchrotron
radiation which corresponds to the break Lorentz factor of electrons,
$\gamma_{\rm br}$, resulting from radiative cooling before escape.
In principle, the break feature may appear in the Compton component, too.
However, the spectral resolution of the present $\gamma$-ray observations
is not good enough.  Moreover, the Klein-Nishina effect makes the
situation complicated.
Thus we do not use the break frequency of the Compton component in this
paper.

\begin{figure} [b!] 
\plotone{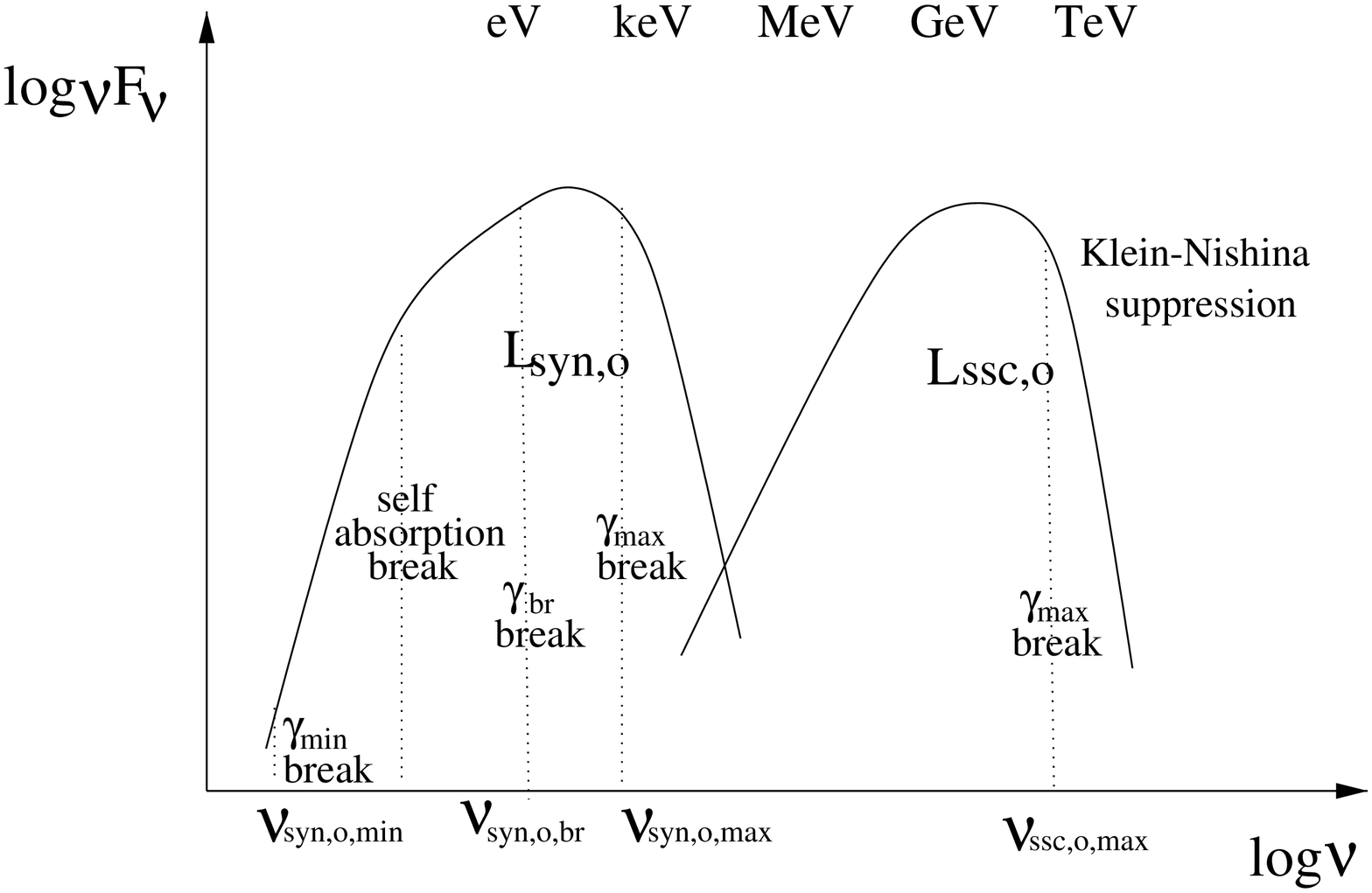}
\caption
{Schematic picture of the multi-frequency spectrum of a typical TeV blazar.
Here $L_{\rm syn,o}$ and $L_{\rm ssc,o}$ are the observed bolometric
luminosities
of synchrotron and SSC components, respectively.
Corresponding to the break in the relativistic electron energy spectrum
in the emission region,
a break feature appears in the observed synchrotron spectrum.
Around the TeV energy region,
the Klein-Nishina effect suppresses the observed flux compared to the
Thomson regime. We exclude the information of
$\nu_{\rm ssc,o,br}$, because of sparse observational data
points and complication from the Klein-Nishina effect.}
\label{fig:photon}
\end{figure}

\begin{figure}  
  \plotone{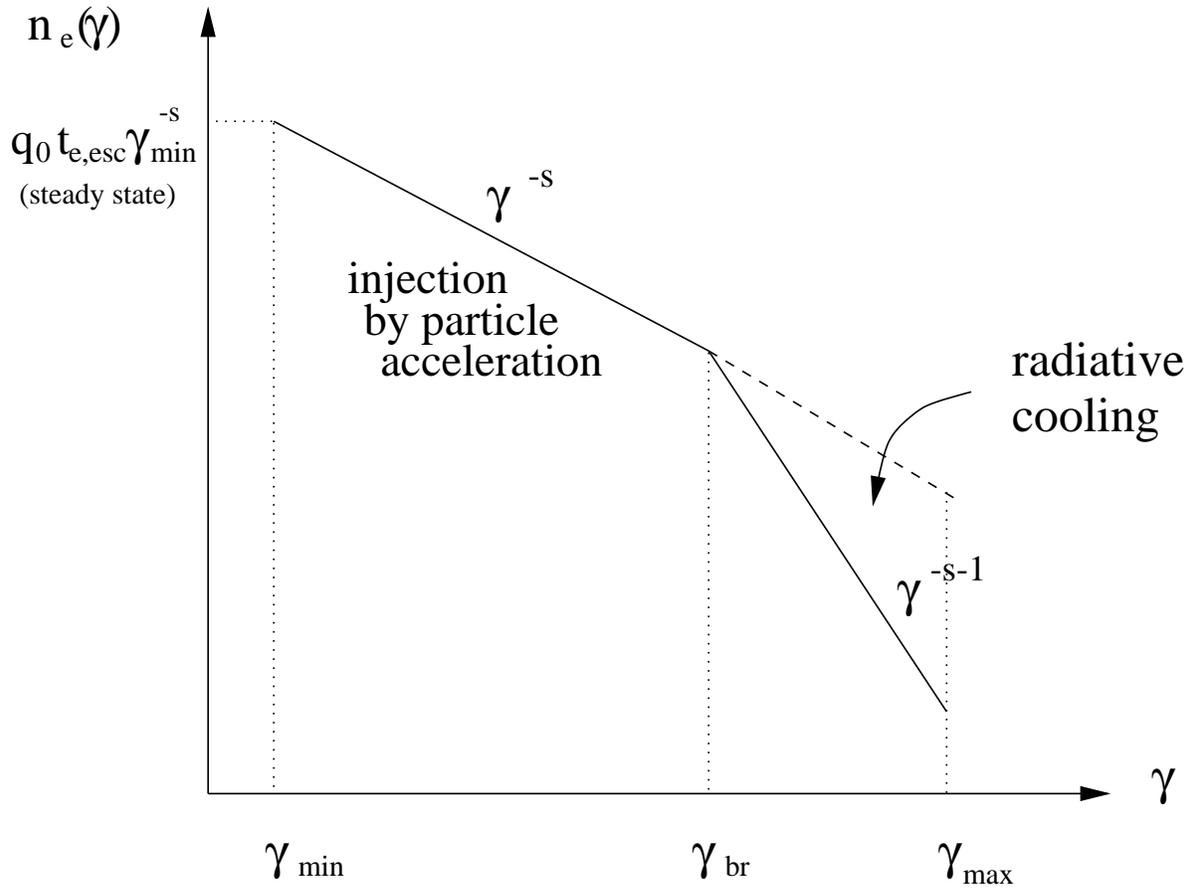}
\caption
{Schematic picture of the relativistic electron energy spectrum.
At high Lorentz factors, radiative cooling
decreases the number density of electrons
and leads to a break in the spectrum.}
\label{fig:electron}
\end{figure}

To sum up, the five typical observables in the observer frame are:
$\nu_{\rm syn,o,max}$ the maximum synchrotron frequency,
$\nu_{\rm syn,o,br}$ the synchrotron break frequency,
$\nu_{\rm ssc,o,max}$ the maximum
frequency of the SSC component,
$L_{\rm syn,o}$ total synchrotron luminosity,
and
$L_{\rm ssc,o}$ total SSC luminosity.
Schematic pictures of a  multi-frequency radiation spectrum
and a relativistic electron energy distribution are shown
in Figures \ref{fig:photon} and \ref{fig:electron}, respectively.
The approximate solution of the electron kinetic equation (\ref{ekin}) is
\begin{eqnarray}\label{slowcool}
n_{\rm e}(\gamma)&=&q_{\rm e} t_{\rm e,esc} \gamma^{-s}
\, \qquad \qquad {\rm for}\quad
 \gamma_{\rm min}\leq \gamma<\gamma_{\rm br}\nonumber \\
n_{\rm e}(\gamma)&=&q_{\rm e} t_{\rm e,esc} \gamma_{\rm br} \gamma^{-s-1}
\qquad {\rm for} \quad \gamma_{\rm br}<\gamma\leq \gamma_{\rm max}.
\end{eqnarray}
provided that $\gamma_{\rm min}< \gamma_{\rm br}$.

Using the standard formula about radiation
(e.g., Rybicki \& Lightman 1979),
we can obtain five relations
between the model parameters and observables.
Observed synchrotron frequencies from a single electron
of Lorentz factor $\gamma_{\rm br}$ and  $\gamma_{\rm max}$ are
respectively given by,
\begin{eqnarray}\label{syn,o,br}
\nu_{\rm syn,o,br} =1.2\times 10^{6}B\gamma_{\rm br}^{2}
\frac{\delta}{1+z}
\end{eqnarray}
and
\begin{eqnarray}\label{syn,o,max}
\nu_{\rm syn,o,max} =1.2\times 10^{6}B\gamma_{\rm max}^{2}
\frac{\delta}{1+z}.
\end{eqnarray}
The maximum value of the observed SSC energy in the Klein-Nishina regime is
\begin{eqnarray}\label{KN}
h\nu_{\rm ssc,o,max} =C_{1}\gamma_{\rm max} m_{\rm e}c^{2}
\frac{\delta}{1+z},
\end{eqnarray}
where $h$ is the Planck constant and
$C_{1}<1$ is a constant representing the uncertainty of the
Klein-Nishina effect which typically taken to be 1/3 here.
It is to be noted that $h \nu_{\rm ssc,o,max}$ is limited by
the Klein-Nishina effect unless the beaming factor is extremely large
(typically larger than about 100). This is understood as follows.
The detection of TeV photons means that $\gamma_{\rm max}$
is at least greater than $10^{6.5}/\delta$.
If $\delta \epsilon_{\rm seed, s}\gamma_{\rm max}^2 \sim 10^{12}$ eV and
$\gamma_{\rm max} > 10^{6.5}/\delta$ are satisfied,
we obtain $\epsilon_{\rm seed, o} < 0.1 \delta^2$ eV,
where $\epsilon_{\rm seed, s}$ and  $\epsilon_{\rm seed, o}$
are seed photon energy in the source frame 
and the observer frame, respectively.
This means that the observed seed photon frequency
is lower than X-ray band unless $\delta >100$.

In the case of  $s< 5/2$, 
bolometric synchrotron luminosity 
in the observer frame is given by
\begin{eqnarray}\label{lumi}
L_{\rm syn,o}=4\pi D^{2}_{\rm L}F_{\rm syn,o}
=\frac{4\pi R^{3}}{3}\delta^{4}
\int^{\gamma_{\rm max}}_{\gamma_{\rm br}}
\frac{4}{3}\sigma_{\rm T}c\gamma^{2}u_{\rm B}q_{\rm e}t_{\rm e,esc}
\gamma_{\rm br}\gamma^{-s-1}
d\gamma,
\end{eqnarray}
where $u_{\rm B}$ is the energy density of magnetic fields,
$D_{\rm L}$ is the luminosity distance, and $F_{\rm syn,o}$
is the total flux of synchrotron radiation in the observer frame.
Note that in the case of $s > 5/2$, 
the luminosity from electrons with
$\gamma_{\rm min}< \gamma<\gamma_{\rm br}$ is 
larger than that from electrons with   
$\gamma_{\rm br}<\gamma<\gamma_{\rm max}$
and equation (\ref{lumi})
is not valid. 
The bolometric Compton luminosity is written in a similar way,
by replacing $u_{\rm B}$ by the energy density of soft photons
and multiplying a suppression factor of $C_{2}$ mentioned below.
The energy density of synchrotron photons is given by
\begin{equation}\label{lsyn}
L_{\rm syn,o}
=\frac{4\pi R^{2}}{3}c\delta^{4}u_{\rm syn} \, ,
\end{equation}
when we set the photon escape time as $R/c$.
It is important to note that because of the Klein-Nishina suppression
only photons with energy less than
$m_{\rm e}c^2/\gamma$ in the source frame
contribute to SSC luminosity.  Here, we simply denote this suppression
factor by $C_2$.
The ratio of the synchrotron and SSC luminosities
is then given by
\begin{eqnarray}\label{lratio}
\frac{L_{\rm syn,o}}{L_{\rm ssc,o}}=\frac{u_{\rm B}}{C_{2}u_{\rm syn}} \, .
\end{eqnarray}
The break Lorentz factor is determined by the condition
that $t_{\rm e,esc}$ is equal to $t_{\rm cool}$:
\begin{eqnarray}\label{tcool}
\frac{R}{C_3 c}=\frac{3m_{\rm e}c}{4(u_{\rm B}+C_{2}u_{\rm syn})\sigma_{\rm
T}
\gamma_{\rm br}} \, ,
\end{eqnarray}
where we set $C_{3}t_{\rm e,esc}=t_{\rm dyn}$
and  assume $C_{3}=1/3$ as has been noted in the previous subsection.
This is a different approach from most of the previous work where
the time variability constraint
$R\sim\delta ct_{\rm var}$ is used.
We think it more appropriate to
avoid time variability constraint in the case of quiescent states,
because the shortest time variability such as 15 minutes TeV flare
(Gaidos et al. 1996)
might be correlated to
local regions such as a shock front 
(Kirk et al. 1998).

\subsubsection{Analytic Estimate of Physical Parameters}\label{parameters}

In this subsection, we analytically estimate the model parameters
using the typical observables  of TeV blazars.
Observed total flux, typical frequencies, and luminosity distance
are scaled as
\begin{eqnarray}
f_{\rm syn}=\frac{F_{\rm syn,o}}{10^{-10}\rm erg \ cm^{-2} \ s^{-1}} \, ,
\quad
f_{\rm ssc}=\frac{F_{\rm ssc,o}}{10^{-10}\rm erg \ cm^{-2} \ s^{-1}} \, ,
\end{eqnarray}
\begin{eqnarray}
\nu_{\rm br}=\frac{\nu_{\rm syn,o,br}}{10^{15}\rm Hz} \, ,\quad
\nu_{\rm max}=\frac{\nu_{\rm syn,o,max}}{10^{17}\rm Hz} \, ,\quad
\nu_{\rm ssc}=\frac{\nu_{\rm ssc,o,max}}{10^{26}\rm Hz} \,\quad
\end{eqnarray}
and
\begin{eqnarray}
d=\frac{D_{\rm L}}{100\ \rm Mpc} \, ,
\end{eqnarray}
respectively.
The numerical factors $C_{1}$, $C_{2}$, and $C_{3}$  normalized
by 1/3 are denoted by $c_1$, $c_2$, and $c_3$, respectively.

In order to express the model parameters 
in terms of the observables,
first we solve
five algebraic equations
(\ref{syn,o,br}),  
(\ref{syn,o,max}),
(\ref{KN}),
(\ref{lratio}), and
(\ref{tcool}), 
to obtain five quantities
$R$,
$B$,
$\delta$,
$\gamma_{\rm br}$, and
$\gamma_{\rm max}$, 
and then by inserting them into Eq. (\ref{lumi}), 
we obtain $q_{\rm e}$. 
As a result, we obtain the typical values of parameters 
for TeV blazars as following:
\begin{eqnarray}
\delta&=&8.9 \
f_{\rm syn}^{1/2}
f_{\rm ssc}^{-1/4}
\eta^{1/2}
\nu_{\rm br}^{1/4}
\nu_{\rm max}^{1/4}
\nu_{\rm ssc}^{-1/2}
d^{1/2}
c_{1}^{1/2}
c_{2}^{1/4}
c_{3}^{-1/2} \, ,
\end{eqnarray}
\begin{eqnarray}
B&=& 0.13\
f_{\rm syn}^{1/2}
f_{\rm ssc}^{-1/4}
\eta^{1/2}
\nu_{\rm br}^{1/4}
\nu_{\rm max}^{5/4}
\nu_{\rm ssc}^{-5/2}
d^{1/2}
(1+z)^{-1}
c_{1}^{5/2}
c_{2}^{1/4}
c_{3}^{-1/2}\ {\rm G} \, ,
\end{eqnarray}
\begin{eqnarray}
\gamma_{\rm br}&=&2.7\times 10^{4}
f_{\rm syn}^{-1/2}
f_{\rm ssc}^{1/4}
\eta^{-1/2}
\nu_{\rm br}^{1/4}
\nu_{\rm max}^{-3/4}
\nu_{\rm ssc}^{3/2}
d^{-1/2}
(1+z)
c_{1}^{-3/2}
c_{2}^{-1/4}
c_{3}^{1/2} \, ,
\end{eqnarray}
\begin{eqnarray}
\gamma_{\rm max}&=&2.7\times 10^{5}
f_{\rm syn}^{-1/2}
f_{\rm ssc}^{1/4}
\eta^{-1/2}
\nu_{\rm br}^{-1/4}
\nu_{\rm max}^{-1/4}
\nu_{\rm ssc}^{3/2}
d^{-1/2}
(1+z)
c_{1}^{-3/2}
c_{2}^{-1/4}
c_{3}^{1/2} \, ,
\end{eqnarray}
\begin{eqnarray}
R&=& 9.0\times 10^{15}
f_{\rm syn}^{-1/2}
f_{\rm ssc}^{1/4}
\eta^{-3/2}
\nu_{\rm br}^{-3/4}
\nu_{\rm max}^{-7/4}
\nu_{\rm ssc}^{7/2}
d^{-1/2}
(1+z)
c_{1}^{-7/2}
c_{2}^{-1/4}
c_{3}^{3/2}\ {\rm cm} \, ,
\end{eqnarray}
and
\begin{eqnarray}
q_{\rm e}\int_{\gamma_{\rm br}}^{\gamma_{\rm max}}
\gamma^{-s+1}d\gamma &=&
0.015\ f_{\rm syn}^{1/2}
f_{\rm ssc}^{1/4}
\eta^{7/2}
\nu_{\rm br}^{5/4}
\nu_{\rm max}^{17/4}
\nu_{\rm ssc}^{-17/2}
d^{3/2}
(1+z)^{-3} \nonumber \\
&& \times c_{1}^{17/2}
c_{2}^{-1/4}
c_{3}^{-5/2}\ {\rm cm}^{-3} \ {\rm s}^{-1} \, ,
\end{eqnarray}
where $\eta= (1+F_{\rm ssc,o}/F_{\rm syn,o})/2$.
Here we emphasize that
these expressions give a complete set of model parameters in terms of
the observables and that typical numerical values turn out to be
similar to those obtained in various other ways
(e.g., Tavecchio et al. 1998; Kataoka et al. 2000).
In particular, the obtained size
seems to be similar to that based on time variability,
which means that the size of the emission region
in quiescent states is compatible with that obtained by time
variability constraint.
It is seen that $R$ and $q_{\rm e}$ have a strong dependence
on $\nu_{\rm max}$ and $\nu_{\rm ssc}$,
and that a variation of $\nu_{\rm ssc}$ by
a factor of 2 leads to a variation of $R$, $B$, and $q_{\rm e}$
by a few orders of magnitude. Despite this, quite robust estimate
is possible for some of the quantities such as the ratio of
the energy density of electrons to that of magnetic fields,
as will be shown in the next section.

Although our method is model dependent in that we use a simple model
of electron injection, escape, and cooling, this model is quite
general and has an advantage of self-consistent treatment of the break
Lorentz
factor of electrons.
If one does not use this relation and tries to
proceed based on observables alone, one needs to introduce
the electron number density instead of the injection parameter
$q_{\rm e}$ and to use time variability constraint.
Most of the previous work adopted such methods and
searched for suitable parameters
in a two dimensional parameters such as ($\log B$, $\log\delta$) plane
(e.g., Tavecchio et al. 1998; Bednarek \& Protheroe 1997; Kataoka et
al. 1999)
by allowing some degree of uncertainties of time variability
constraint.

Next we check an additional constraint on the transparency
of $\gamma$-rays
against the intrinsic absorption,
i.e., the optical depth of a $\gamma$-ray photon for pair production
should be smaller than unity.
For a $\gamma$-ray photon of the observed energy
$\epsilon_{\rm \gamma,o}$,
the observed energy of the target photon $\epsilon_{\rm t,o}$
is about
\begin{eqnarray}
\epsilon_{\rm t,o}=\frac{\delta}{1+z}\epsilon_{\rm t,s}
=\frac{4m_{\rm e}^2c^4}{\epsilon_{\rm \gamma,o}}
      (\frac{\delta}{1+z})^2 \, .
\end{eqnarray}
We approximate the number density of target photons by
\begin{eqnarray}
n_{\rm t,s}=\frac{3L_{\rm t,s}}{4\pi R^2 c \epsilon_{\rm t,s}}.
\end{eqnarray}
Then, the optical depth is given by 
(von Montigny et al. 1995):
\begin{eqnarray}
\tau_{\gamma \gamma}(\epsilon_\gamma)
&=& \frac{15}{256\pi}\frac{1+z}{\delta^5}
\frac{\epsilon_{\rm \gamma,o} L_{\rm t,o}\sigma_{\rm T}}{R m_{\rm e}^2 c^5}
\nonumber \\
&=& 9.9 \times 10^{-2}
\frac{ L_{\rm t,o}}{10^{44}{\rm erg \ s^{-1}}}
\frac{\epsilon_{\rm \gamma,o}}{\rm TeV}
\left(\frac{R}{10^{16}{\rm cm}}\right)^{-1}
(\frac{\delta}{10})^{-5}(1+z).
\end{eqnarray}
If we use the typical observables, the optical depth is expressed as
\begin{eqnarray}
\tau_{\gamma \gamma}(\epsilon_\gamma)
&=& 9.8\times 10^{-2}\ \frac{L_{\rm t,o}}{L_{\rm syn,o}}
f_{\rm syn}^{-1}
f_{\rm ssc}
\eta^{-1}
\nu_{\rm br}^{-1/2}
\nu_{\rm max}^{1/2}
(1+z)^{-5}
c_{1}
c_{2}^{-1}
c_{3}
\frac{\epsilon_{\rm \gamma,o}}{h\nu_{\rm ssc,o,max}} \, .
\end{eqnarray}
Thus we find that intrinsic absorption of 0.35 TeV photons is not
so large for TeV blazars in quiescent states
but it may be important for some cases.
In \S \ref{result}, we will show the value
for numerical solutions in Tables 1, 2, and 3.

\section{Energy Density of Relativistic Electrons and Magnetic Field}
\label{energy}

In this section, following the methods of
\cite{t97},
we show that we can analytically estimate the ratio of $u_{\rm B}$
to $u_{\rm e}$ fairly robustly.
The energy density of relativistic electrons is given by
\begin{eqnarray}\label{ue}
u_{\rm e}&=& \int_{\gamma_{\rm min}}^{\gamma_{\rm max}}
          \gamma m_{\rm e} c^{2}n_{\rm e}(\gamma) d\gamma \nonumber \\
&\simeq& q_{\rm e} t_{\rm e,esc}m_{\rm e}c^{2}
\int_{\gamma_{\rm min}}^{\gamma_{\rm br}} \gamma^{-s+1}d\gamma \nonumber \\
&=& 1.1 \times 10^{-2}
f_{\rm ssc}^{1/2}
\eta^{2}
\nu_{\rm br}^{1/2}
\nu_{\rm max}^{5/2}
\nu_{\rm ssc}^{-5}
d
(1+z)^{-2}
c_{1}^{5}
c_{2}^{-1/2}
c_{3}^{-2} \times
\frac{\int_{\gamma_{\rm min}}^{\gamma_{\rm br}} \gamma^{-s+1}d\gamma}
{\int_{\gamma_{\rm br}}^{\gamma_{\rm max}} \gamma^{-s+1}d\gamma}
\quad {\rm erg \ cm^{-3}} \, ,
\end{eqnarray}
in the framework of the previous section.
The energy density of magnetic fields is also given by
\begin{eqnarray}
u_{\rm B}&=&6.3 \times 10^{-4}
f_{\rm syn}
f_{\rm ssc}^{-1/2}
\eta
\nu_{\rm br}^{1/2}
\nu_{\rm max}^{5/2}
\nu_{\rm ssc}^{-5}
d
(1+z)^{-2}
c_{1}^{5}
c_{2}^{1/2}
c_{3}^{-1} \quad {\rm erg \ cm^{-3}} \, .
\end{eqnarray}
Although both energy densities have fairly strong dependence
on some observables,
the ratio turns out to be as simple as
\begin{eqnarray} \label{ratio}
\frac{u_{\rm e}}{u_{\rm B}}  &=&
9\ {c_2}^{-1}{c_3}^{-1}
\left( 1 + \frac{f_{\rm ssc}}{f_{\rm syn}} \right )
\frac{f_{\rm ssc}}{f_{\rm syn}}
\frac{\int_{\gamma_{\rm min}}^{\gamma_{\rm br}} \gamma^{-s+1}d\gamma}
{\int_{\gamma_{\rm br}}^{\gamma_{\rm max}} \gamma^{-s+1}d\gamma} \, .
\end{eqnarray}
Note that the strong dependence on quantities such as $\nu_{\rm ssc}$,
$\nu_{\rm max}$, and $c_1$ in $u_{\rm e}$ and $u_{\rm B}$ are
canceled out.
In addition, we can derive equation (\ref{ratio})
in a more transparent way as follows.
The total radiation power is simply given by
\begin{eqnarray}\label{ltot}
L_{\rm syn,o}+L_{\rm ssc,o}
&=&\frac{4}{3}\pi R^{3}\delta^{4}m_{\rm e}c^2
q_{\rm e}\int^{\gamma_{\rm max}}_{\gamma_{\rm br}} \gamma^{-s+1} d\gamma.
\end{eqnarray}
Combining equations
(\ref{lsyn}), (\ref{lratio}), (\ref{ue}), and (\ref{ltot})
we can easily obtain equation (\ref{ratio})
(The factor $9$ in Eq.(\ref{ratio}) 
comes from the normalization of $C_{2}$ and $C_{3}$.).
This analytic estimation is quite useful to understand the relation
between the typical observables of TeV blazars and the ratio of
$u_{\rm e}/u_{\rm B}$.
Needless to say, the ratio of $\gamma_{\rm br}$
to $\gamma_{\rm max}$ is obtained from the observed ratio of
$\nu_{\rm syn,o,br}$ to $\nu_{\rm syn,o,max}$
(Eqs. \ref{syn,o,br} and \ref{syn,o,max}). The ratio of $\gamma_{\rm br}$
to $\gamma_{\rm min}$ depends on the adopted value of $\gamma_{\rm min}$.
When the synchrotron luminosity dominates over the SSC luminosity,
$u_{\rm syn}\ll u_{\rm B}$ and $u_{\rm e}/u_{\rm B}
\propto f_{\rm ssc}/f_{\rm syn} \ll 1$.
When the SSC luminosity dominates over the synchrotron luminosity,
$u_{\rm syn}\gg u_{\rm B}$  and $u_{\rm e}/u_{\rm B} \propto
f_{\rm ssc}^2/f_{\rm syn}^2 \gg 1$.
Hence, as the value of $f_{\rm ssc}/f_{\rm syn}$ increases,
the value of $u_{\rm e}/u_{\rm B}$ increases.
Very roughly, the equi-partition between electrons and magnetic fields
corresponds to sources for the SSC luminosity equal to
the synchrotron luminosity, if we ignore other numerical factors
such as $c_2$ and $c_3$.
According to equation (\ref{ratio}),
more realistic estimate indicates that
when the SSC luminosity is equal to
the synchrotron one,  $u_{\rm e}/u_{\rm B}$
takes a value of an order of 10.

Since the above estimate depends on several numerical factors,
we should examine carefully these subtleties.
As for the value of $C_2$, the energy of seed photons for $\gamma$-rays
of $\epsilon_{\rm \gamma,o}$ TeV is lower than about
$10\epsilon_{\rm \gamma,o}^{-1}$ eV for a typical value of $\delta = 10$.
Because of the Klein-Nishina effects, synchrotron photons whose energy is
higher than this value is not available to inverse Compton scattering.
Thus $C_2$ is expected to be less than 0.3.
As for $C_3$, since electron escape 
may be identified with
the expansion velocity
in the downstream region of the shock, 1/3 is also a reasonable guess;
higher values approaching 1 requires as rapid escape of electrons as that of
photons, which seems to be difficult to realize.
The typical ratio of $\gamma_{\rm max}$ to $\gamma_{\rm br}$ is 10.
Even if this ratio is 100, $u_{\rm e}/u_{\rm B}$ decreases only by a factor
of a few when $s=2$.  An increase in the value of $\gamma_{\rm min}$
by a factor of 10 will
decrease $u_{\rm e}/u_{\rm B}$ by a factor of a few for $s=2$.
Thus, only when both $\gamma_{\rm min}$ and
$\gamma_{\rm max}/\gamma_{\rm br}$
are larger than 100, $u_{\rm B}$ can be comparable to $u_{\rm e}$.
We should note that since $\gamma_{\rm min}$ is expected to
be comparable with the value of the bulk Lorentz factor
in the shock acceleration picture, $\gamma_{\rm min} = 10$ 
is a reasonable choice 
when the jet material consists of mainly electron-positron pairs
or when electrons and ions are separately thermalized for normal plasmas
(see \S \ref{dis}, for further discussion).

Thus, the only major uncertainty in estimating $u_{\rm B}/u_{\rm e}$
is the electron index $s$.
In a previous work of
\cite{t97},
he examined the case of $s = 2$, because this is the universal index
expected
for non-relativistic case
(e.g., Blandford \& Eichler 1987).
In the present work, we allow this value to be less than 2,
which better fits the emission spectra.

\begin{figure}  
  \plotone{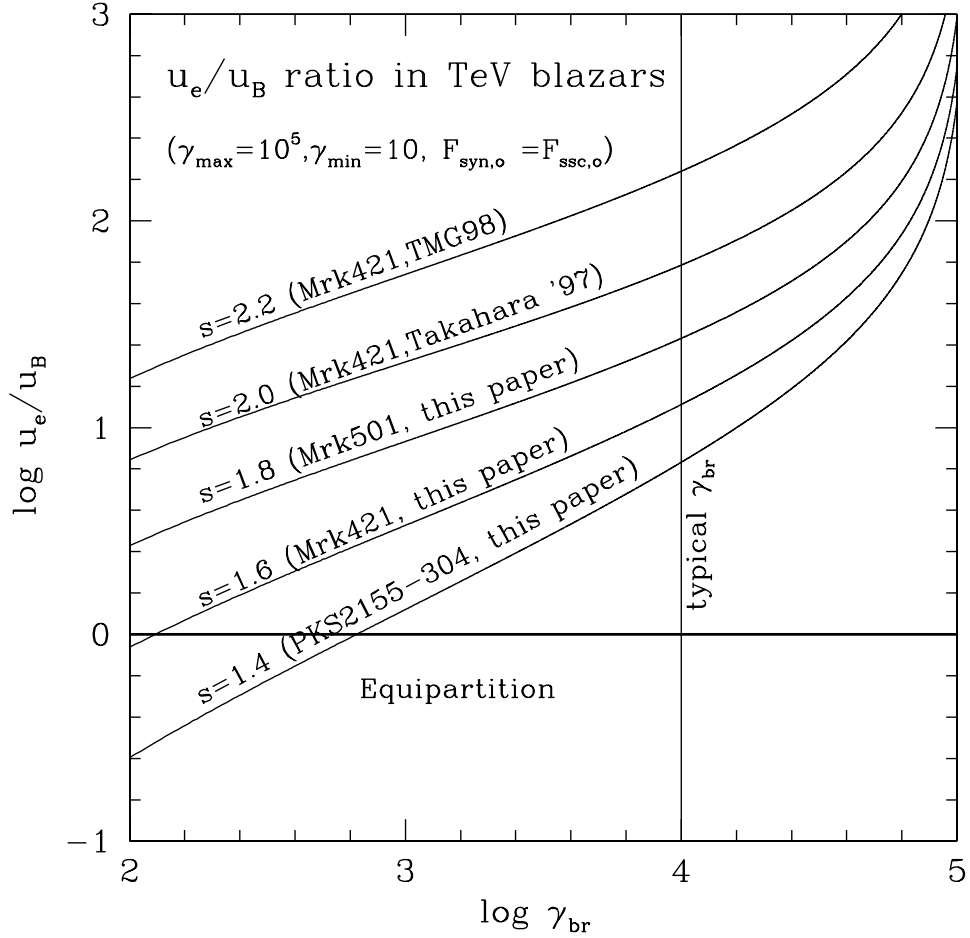}
\caption
{Ratio $u_{\rm e}/u_{\rm B}$ in TeV blazars.  This is calculated
according to equation (\ref{ratio}).
For a TeV blazar $u_{\rm e}$ is typically about one order of magnitude
larger than $u_{\rm B}$.
We confirm this analytic result by numerical
calculations.
The discrepancies
between this analytic estimation and numerical results
are at most within a factor of a few.}
\label{fig:ueub}
\end{figure}

When $s$ is smaller, $u_{\rm e}/u_{\rm B}$ becomes smaller.
In Figure \ref{fig:ueub} we show
the resultant ratio $u_{\rm e}/u_{\rm B}$ as a function
of $\gamma_{\rm br}$ for several values of $s$.
For simplicity, other parameters are fixed at
$f_{\rm syn}=1,
 f_{\rm ssc}=1,
 \gamma_{\rm max}= 1\times10^{5},
 \gamma_{\rm min}= 10,
 c_{1}= 1,
 c_{2}= 1$,
and $c_{3}=1$.
Figure \ref{fig:ueub} shows that even when $s$ is as small as 1.4,
$u_{\rm e}$ is larger than $u_{\rm B}$,
unless $\gamma_{\rm br}$ is smaller than $10^3$.
Thus, the conclusion that $u_{\rm e}$ is about one order of magnitude
larger than $u_{\rm B}$ is fairly robust.

Before we present detailed numerical results, we describe straightforward
relations between the energy densities and power carried by relativistic
jets.
It should be noted that the energy density of $i$-th radiation component
(the suffix $i$ corresponds to synchrotron or SSC),
$u_{i}$
is related to the power $L_{i}$ as
\begin{eqnarray}\label{li}
L_{i}
=\frac{4}{3}\pi R^{2}cu_{i}\Gamma^{2} \, ,
\end{eqnarray}
where the factor $4/3$ accounts for pressure of relativistic matter and
we assume $\Gamma \gg 1$.
It is worth noting that the
observed luminosity is given by
\begin{eqnarray}\label{lio}
L_{\rm i,o}=\frac{4}{3}\pi R^{2}cu_{i}\delta^{4}
\end{eqnarray}
in the observer frame.
Hence the actual radiation power is smaller than the observed luminosity
by a factor of $\delta^{2}$ assuming $\Gamma\sim\delta$.
Similarly, the Poynting power is given by
\begin{equation}
L_{\rm Poy} =
\frac{4}{3}\pi R^{2}c u_{\rm B} \Gamma^{2}
\end{equation}
and the kinetic power of relativistic electrons is
\begin{equation}
L_{\rm e,kin}
= \frac{4}{3}\pi R^{2} c u_{\rm e} \Gamma^{2} \, .
\end{equation}
Thus, the discussion on the ratio $u_{\rm e}/u_{\rm B}$ is
straightforwardly translated into the ratio $L_{\rm e,kin}/L_{\rm Poy}$.
It should be stressed that $L_{\rm e, kin}$
takes account of only relativistic electrons.
In other words,
the contribution from thermal electrons which should
constitute a reservoir for
acceleration and that of protons, either relativistic or cold,
are completely neglected.
We emphasize that,
if we take into account of these components,
it is clear that kinetic power is more dominant
at least in the emission region of TeV blazars.

\section{NUMERICAL RESULTS}\label{result}

Here, we present numerical results by solving the kinetic equations
and searching for best fitted model parameters for three TeV blazars
(Mrk 421, Mrk 501, and PKS 2155--304).
The search is made around the parameters determined analytically
in the previous section.  The observed values are taken from
those compiled by \cite{k99}.
In his paper,
the multi-frequency observed spectrum is fitted with a polynomial function
of the form of
$\log(\nu L_{\nu})=a+b\log\nu+c(\log\nu)^{2}+d(\log\nu)^{3}$,
where $a$, $b$, $c$, and $d$ are the fitting constants
(Comastri, Molendi, \& Ghisellini 1995).
\cite{k99}
identified $\nu_{\rm syn,o,br}$ and $\nu_{\rm syn,o,max}$
as the frequencies where synchrotron luminosity reaches a half of its
peak value; the lower one is identified as $\nu_{\rm syn,o,br}$
and the higher one is identified as $\nu_{\rm syn,o,max}$.
In the same way, $\nu_{\rm ssc,o,max}$ can be defined.
However, since the spectra of high energy $\gamma$-rays are not so well
covered, we do not put a heavy weight on his determination of
$\nu_{\rm ssc,o,max}$, but we simply assume
$\nu_{\rm ssc,o,max}=1\times10^{26}\rm Hz$ as a starting point.
It should be noted that these observables are not derived from the spectral
fitting described in the previous section,
but from the polynomial fitting, so that the chosen values of the model
parameters
above are expected to deviate from our definition by some factor.

Moreover, taking
the sparseness and uncertainties of
the observed data into consideration,
it is natural to think some range of uncertainties are included
in the model parameters, too.
Such uncertainties are also investigated in this section.
Here, we explain the procedure of searching these parameter sets.
The range of uncertainties is different for each model parameter.
Among seven model parameters, the index $s$ can be regarded
as fixed. In contrast, $R$ and $q_{\rm e}$ are expected to have
a large uncertainty, while $\delta$ is relatively stable.
The search for model parameters is made as follows.
We first choose a certain value of $q_{\rm e}$.
Then, we adopt a suitable value of $R$ so as to reproduce
the low energy spectrum of the synchrotron component, whereby
some discrepancies will appear at the high energy part of the
synchrotron component. Noting that the observed ratio of
$L_{\rm ssc,o}/L_{\rm syn,o}$ is determined by the combination
of $B^2 \delta^4 R^2$, we can adjust the break feature of
the synchrotron spectrum by adopting a suitable combination of
$B$ and $\delta$. A slight adjustment of the value of $R$
is also made. Finally, the high energy end of the synchrotron
component is adjusted by adopting a suitable $\gamma_{\max}$.
In this step, slight adjustment of the model parameters
which are determined in the previous steps is also made.
The resultant spectral shape of the SSC component should be compared
with observation, while the calculated SSC luminosity should
match the observation. This comparison determines whether the
chosen parameter set is allowable or not.
The whole step is repeated starting from a different value of $q_{\rm e}$
to find the best fit model parameters and uncertainties.
Although this is not a complete survey of the parameter space,
we believe that this provides a reasonable estimate of the
uncertainties because the most uncertain parameters are $q_{\rm e}$ and $R$.

\subsection{Best Fit Parameters}
\subsubsection{Mrk 421}

Mrk 421 ($z = 0.031$) is a well known BL Lac object and the first
identified source of TeV gamma-ray emission by
Whipple Cherenkov telescope 
(Punch et al. 1992).

\begin{figure}   
  \plotone{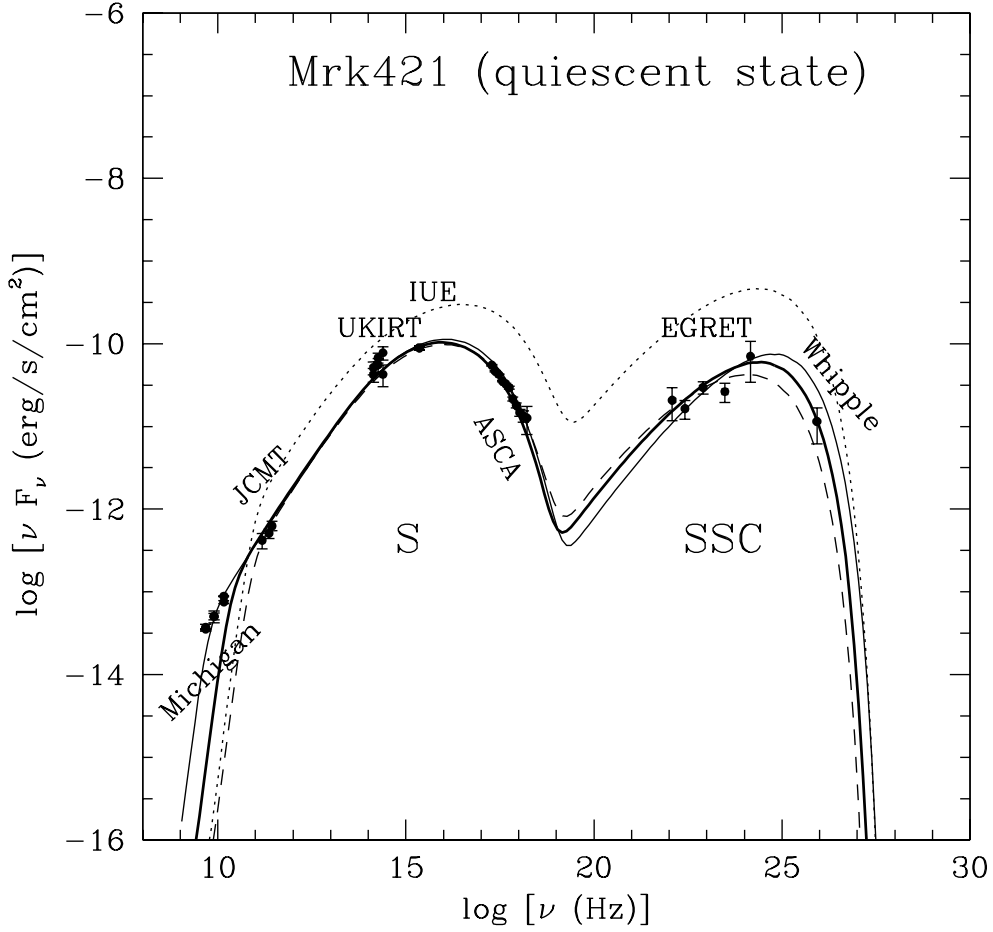}
\caption
{One-zone SSC model spectra for the steady state emission of Mrk 421.
The thick solid line shows the best fit spectrum, 
where adopted parameters are
$\delta = 12$,
$R = 2.8 \times 10^{16}$ cm,
$B = 0.12$ G,
$\gamma_{\rm max}=1.5\times 10^{5}$,
$q_{\rm e} = 9.6 \times 10^{-6}$ cm$^{-3}$ s$^{-1}$,
$s = 1.6$, and
$u_{\rm e}/u_{\rm B} = 5$.
The dotted line shows the spectrum obtained
using the analytic estimates for Mrk 421.
The thin solid and dashed lines show the spectra
of low and high injection models, respectively,
to indicate the uncertainty range
of the spectral fitting.}
\label{fig:mrk421}
\end{figure}

Following
\cite{k99}, the observables of Mrk 421 are chosen as follows;
$\alpha=0.3$,
$\nu_{\rm syn,o,br}=2.5\times10^{14}$ Hz,
$\nu_{\rm syn,o,max}=1.6\times 10^{17}$ Hz,
$\nu_{\rm ssc,o,max}= 1.0\times 10^{26}$ Hz,
$F_{\rm syn,o}=8.6\times 10^{-10}$ erg cm$^{-2}$ s$^{-1}$,
and $F_{\rm ssc,o}=3.4\times 10^{-10}$ erg cm$^{-2}$ s$^{-1}$.
Substituting these observables into the analytic estimate
described in \S 2, we obtain
$\delta = 14.3$,
$R = 7.9\times 10^{15}$ cm,
$B = 0.31$ G,
$\gamma_{\rm br} = 0.69 \times 10^{4}$,
$\gamma_{\rm max} = 1.8 \times 10^{5}$,
$q_{\rm e}=3.2\times 10^{-4}$ cm$^{-3}$ s$^{-1}$, and
$s = 1.6$.  Dotted line in Figure \ref{fig:mrk421} shows
the predicted spectrum obtained by using these analytic values.

This first set of parameters in fact produces much higher luminosities
than observations, which is not surprising as explained above.
Then we try to search for the best fit parameters by changing
model parameters to match the predictions with observations.
Finally, we find a satisfactory set of model parameters as
$\delta = 12$,
$R = 2.8\times 10^{16}$ cm,
$B = 0.12$ G,
$\gamma_{\rm max} = 1.5 \times 10^{5}$,
$q_{\rm e} = 9.6 \times 10^{-6}\rm\ cm^{-3}\ s^{-1}$,
and $s=1.6$.
In this fitting procedure, we fix the reference value of $s$,
because $s$ can be determined with little error from
the spectral shape of the low energy synchrotron emission.
The thick solid curve in Figure \ref{fig:mrk421}
shows the predicted spectrum of Mrk 421
calculated from equations (\ref{ekin}) and (\ref{pkin})
for the best parameter set given above.
The corresponding electron energy spectrum is shown by the
thick solid curve in Figure 5.
The ratio of $u_{\rm e}/u_{\rm B}$
derived using this self-consistent numerical result is
$u_{\rm e}/u_{\rm B} = 5$.

\begin{figure}   
  \plotone{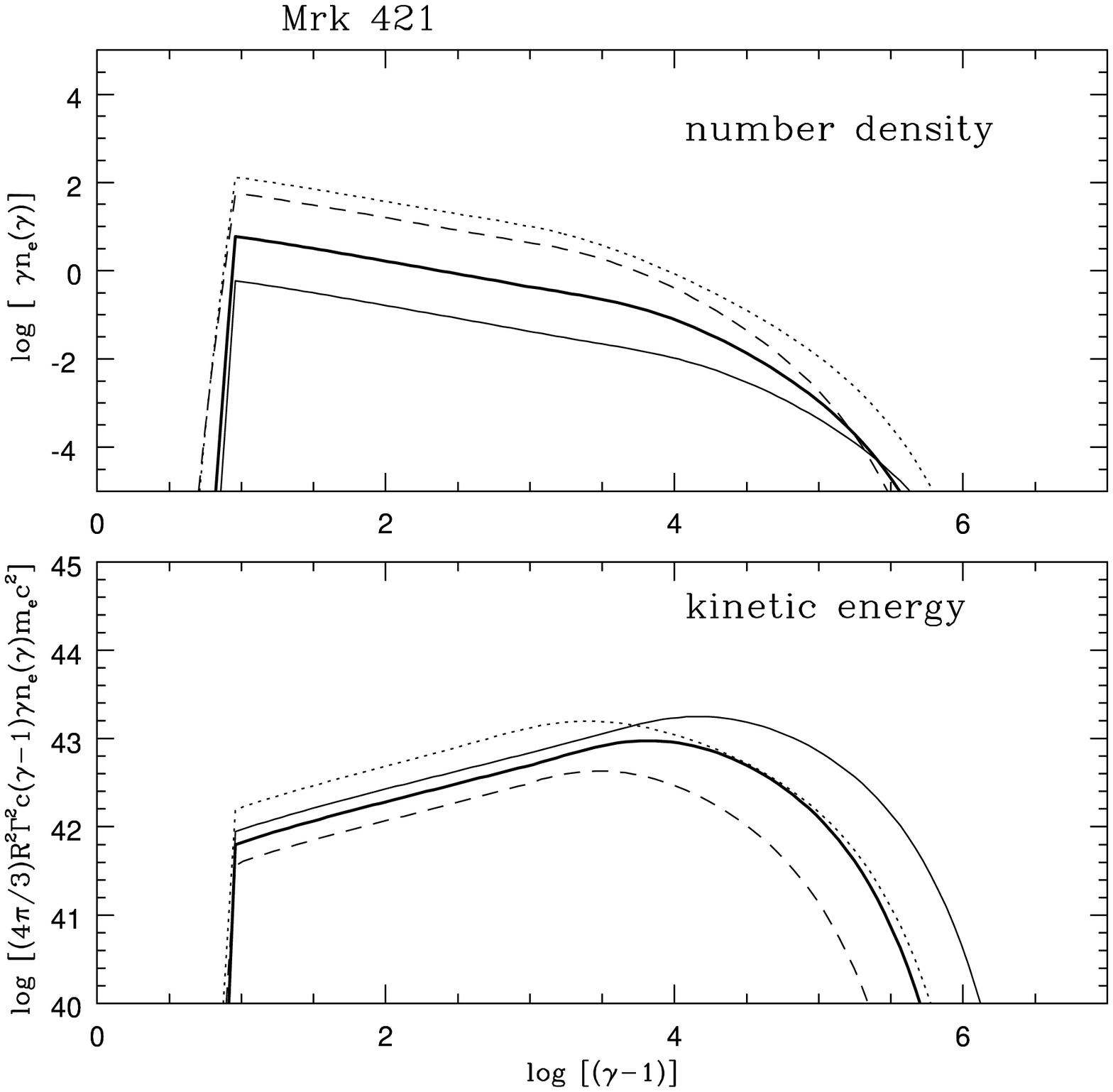}
\caption
{Electron energy spectrum and kinetic power
of Mrk 421 corresponding to Figure \ref{fig:mrk421}.}
\label{fig:mrk421tot}
\end{figure}

The thin solid  curve and the dashed curve
in Figures \ref{fig:mrk421} and \ref{fig:mrk421tot}
are for different sets of
model parameters to examine the range of uncertainties in the values
of the model parameters.
The dashed curve is for the case that $q_{\rm e}$
is 50 times larger than the best fitted value.
The thin solid curve is for the case that $q_{\rm e}$ is 50 times smaller
than the best fit value.
These two sets of parameters may be regarded as marginally allowed,
although the discrepancy in the TeV range is fairly large.
The resultant values of the model parameters are tabulated in Table 1.
As is seen, for the larger value of $q_{\rm e}$, the values of
$R$ and $\gamma_{\rm max}$ become smaller,
while the values of $B$ and $\delta$ become larger.
For a smaller value of $q_{\rm e}$, the reverse is true.
Table 1 shows
the range of uncertainties for model parameters too.
In particular, the ratio of
$u_{\rm e}/u_{\rm B}$ is uncertain by a factor of a few and
the dominance of electrons over magnetic field is not changed.
The values of $\delta$, $\gamma_{\rm max}$, and $B$ are
uncertain by factors of about 2, 5, and 10, respectively.
The size of the emission region is least constrained with
an uncertainty by a factor of 30, but covers a reasonable range.
These results can be roughly understood by combining
equations (\ref{syn,o,br}), (\ref{syn,o,max}),
(\ref{lumi}), and (\ref{lratio}).

\subsubsection{Mrk 501}

BL Lac object Mrk 501 ($z = 0.034$) is also a well known source of
TeV $\gamma$-rays detected by Whipple Cherenkov telescope
(Quinn et al. 1996).
Observables of Mrk 501 are chosen as follows 
(Kataoka 2000);
$\alpha = 0.4$,
$\nu_{\rm syn,o,br} = 6.3\times10^{13}$ Hz,
$\nu_{\rm syn,o,max} = 4.0\times 10^{17}$ Hz,
$\nu_{\rm ssc,o,max} = 1.0\times 10^{26}$ Hz,
$F_{\rm syn,o}=3.5\times 10^{-10}$ erg cm$^{-2}$ s$^{-1}$, and
$F_{\rm ssc,o}=4.5\times 10^{-10}$ erg cm$^{-2}$ s$^{-1}$.
Substituting these values into the analytic estimate,
we obtain
$\delta = 10.1$,
$R = 3.5 \times 10^{15}$ cm,
$B=0.55$ G,
$\gamma_{\rm br} = 0.31\times 10^{4}$,
$\gamma_{\rm max} = 2.5\times 10^{5}$,
$q_{\rm e} = 2.4 \times 10^{-2}$ cm$^{-3}$ s$^{-1}$,
and $s = 1.8$.

\begin{figure}   
  \plotone{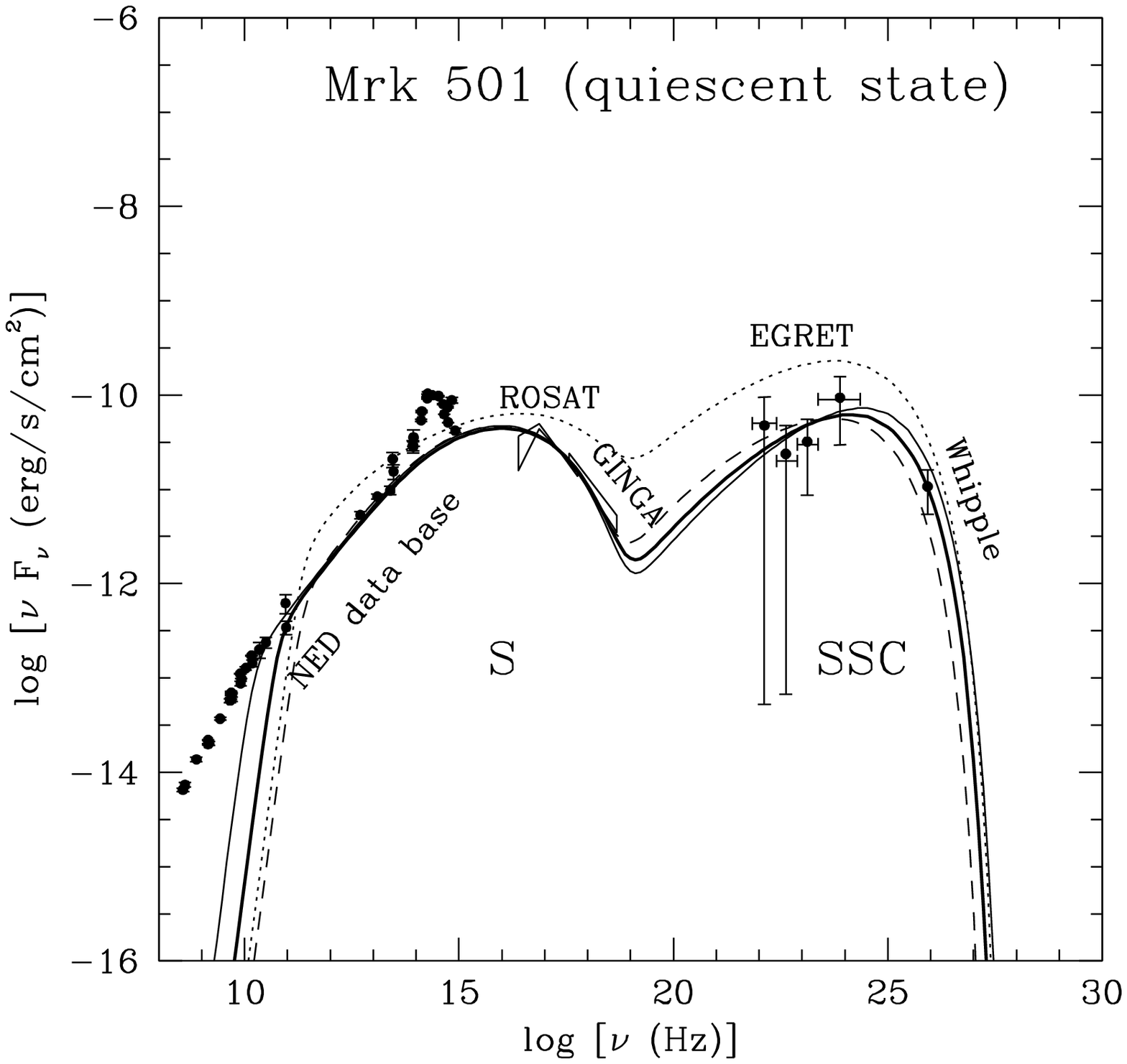}
\caption
{One-zone SSC model spectra for the steady state emission of Mrk 501.
The thick solid line shows the best fit spectrum where
adopted parameters are
$\delta = 11$,
$R = 1.0 \times 10^{16}$ cm,
$B = 0.20$ G,
$\gamma_{\rm max} = 2.0 \times 10^{5}$,
$q_{\rm e} = 1.7 \times 10^{-3}$ cm$^{-3}$ s$^{-1}$,
$s = 1.8$, and
$u_{\rm e}/u_{\rm B} = 22$.
The dotted line shows the spectrum obtained
using the analytic estimates for Mrk 501.
The thin solid and dashed lines show the spectra
of low and high injection models, respectively,
to indicate the uncertainty range
of the spectral fitting.}
\label{fig:mrk501}
\end{figure}

\begin{figure}   
  \plotone{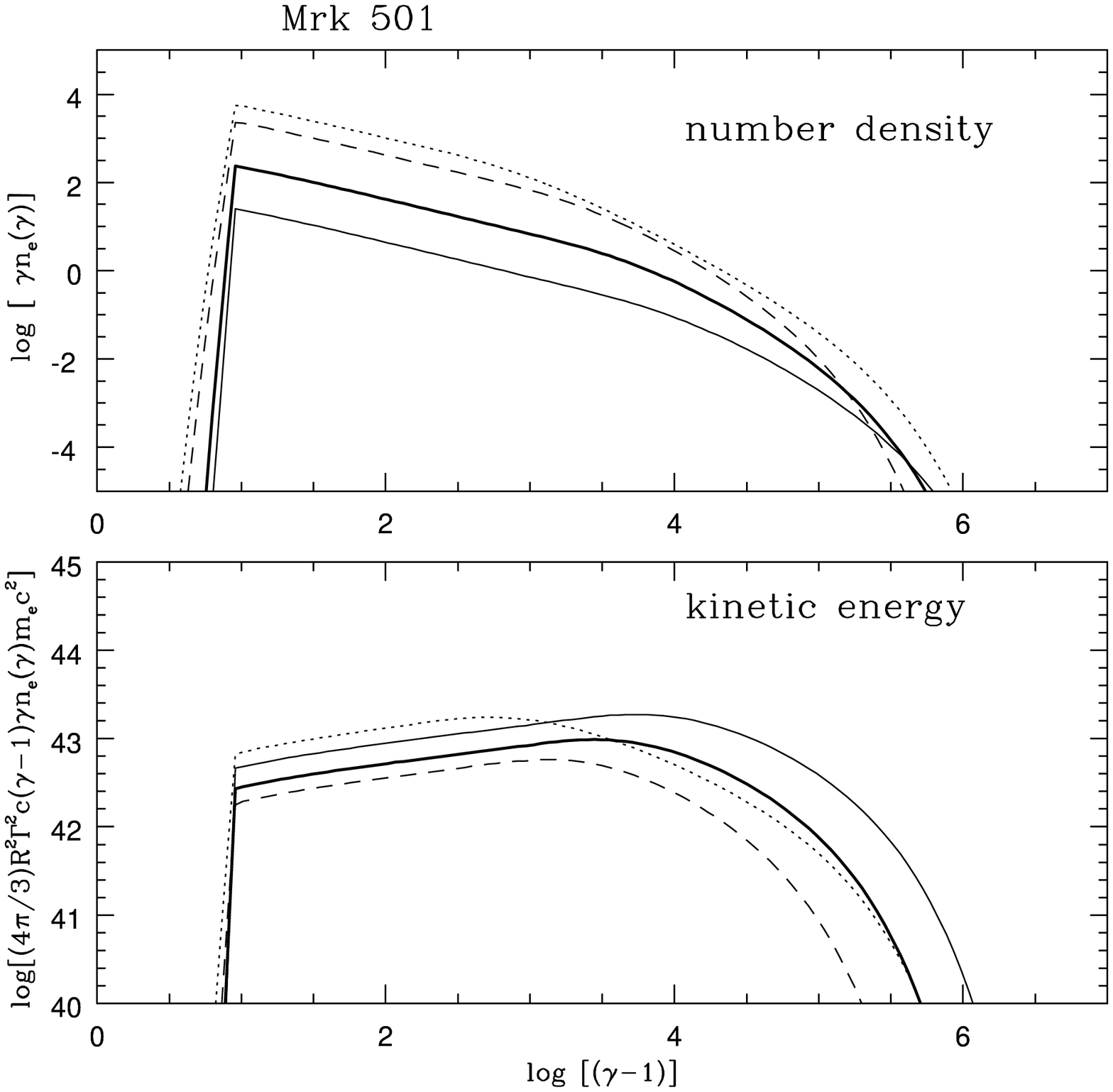}
\caption
{Electron energy spectrum and kinetic power
of Mrk 501 corresponding to Figure \ref{fig:mrk501}.}
\label{fig:mrk501tot}
\end{figure}

In a similar way to the case of Mrk 421, we find that
a satisfactory fit is obtained for the parameters as
$\delta = 11$,
$R = 1.0 \times 10^{16}$ cm,
$B = 0.20$ G,
$\gamma_{\rm max} = 2.0\times 10^{5}$,
$q_{\rm e}=1.7\times 10^{-3}$ cm$^{-3}$ s$^{-1}$,
and $s = 1.8$.
The thick solid curve in Figure \ref{fig:mrk501} shows
the predicted spectrum of
Mrk 501 obtained by numerically calculating
equations (\ref{ekin}) and (\ref{pkin})
for the best fit parameter set given above.
The corresponding electron energy spectrum is shown by thick solid curves
in Figure \ref{fig:mrk501tot}.
The ratio of $u_{\rm B}/u_{\rm e}$
derived from this numerical result is
$u_{\rm e}/u_{\rm B} = 22$.

As is for Mrk 421, the model predictions for the injection rates 50 times
higher and lower than the best fit value
are depicted by dashed and thin solid curves in Figures \ref{fig:mrk501}
and \ref{fig:mrk501tot}.
The numerical values are tabulated in Table 2.
Since the trend of uncertainties is the same as for Mrk 421,
we do not repeat it here.

\subsubsection{PKS 2155--304}

TeV emission from PKS 2155--304 ($z = 0.117$) was detected very recently by
Durham Mark 6 Cherenkov telescope 
(Chadwick et al. 1999).
Observables of PKS 2155--304 are chosen as follows 
(Kataoka 2000);
$\alpha = 0.2$,
$\nu_{\rm syn,o,br} = 4.0 \times10^{14}$ Hz,
$\nu_{\rm syn,o,max} = 2.0 \times 10^{17}$ Hz,
$\nu_{\rm ssc,o,max} = 1.0 \times 10^{26}$ Hz,
$F_{\rm syn,o} = 1.2 \times 10^{-9}$ erg cm$^{-2}$ s$^{-1}$, and
$F_{\rm ssc,o} = 5.8 \times 10^{-10}$ erg cm$^{-2}$ s$^{-1}$.
Substituting these observables into the analytic estimate,
we obtain
$\delta = 35.5$,
$R = 1.9 \times 10^{15}$ cm,
$B = 0.89$ G,
$\gamma_{\rm br} = 0.34 \times 10^{4}$,
$\gamma_{\rm max} = 0.76 \times 10^{5}$,
$q_{\rm e} = 2.3 \times 10^{-3}$ cm$^{-3}$ s$^{-1}$,
and $s = 1.4$.

\begin{figure}   
  \plotone{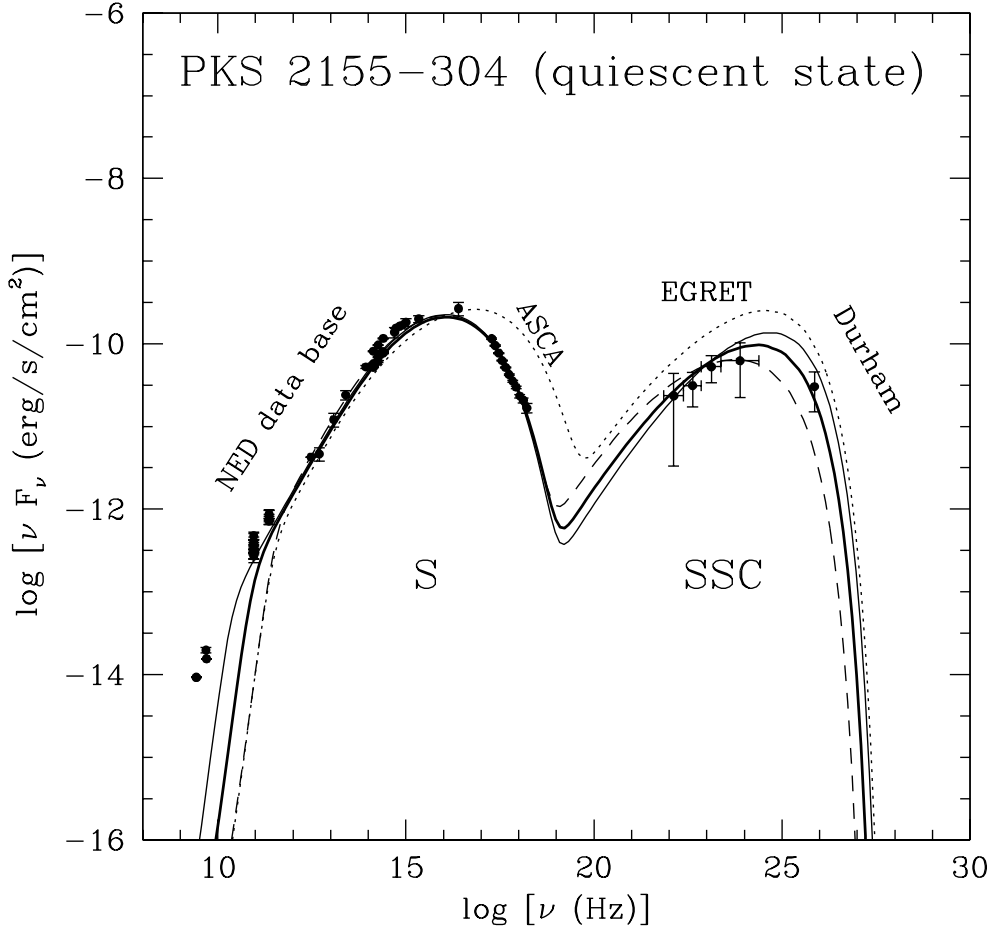}
\caption
{One-zone SSC model spectra for the steady state emission of PKS 2155--304.
The thick solid line shows the best fit spectrum where
adopted parameters are
$\delta = 33$,
$R = 9.0 \times 10^{15}$ cm,
$B = 0.30$ G,
$\gamma_{\rm max} = 0.5 \times 10^{5}$,
$q_{\rm e} = 2.8 \times 10^{-5}$ cm$^{-3}$ s$^{-1}$,
$s = 1.4$,
and $u_{\rm e}/u_{\rm B} = 3$.
The dotted line shows the spectrum obtained
using the analytic estimates for PKS 2155--304.
The thin solid and dashed lines show the spectra
of low and high injection models, respectively,
to indicate the uncertainty range
of the spectral fitting.}
\label{fig:pks2155}
\end{figure}

\begin{figure}   
  \plotone{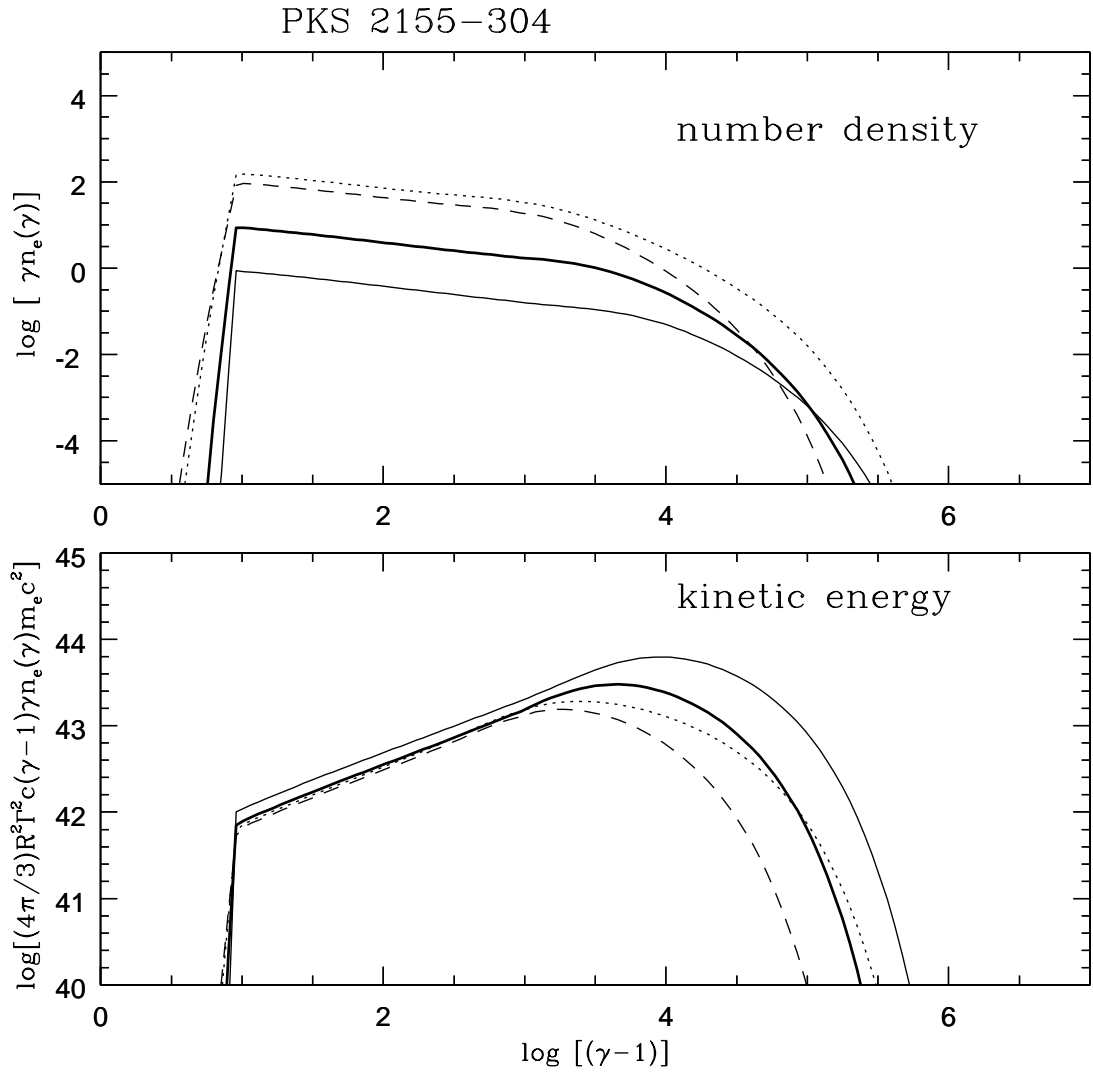}
\caption
{Electron energy spectrum and kinetic power
of PKS 2155--304 corresponding to Figure \ref{fig:pks2155}.}
\label{fig:pks2155tot}
\end{figure}

In a similar way to the cases of Mrk 421 and Mrk 501, we find that
a satisfactory fit is obtained for the parameter set given by
$\delta = 33$,
$R = 9.0 \times 10^{15}$ cm,
$B = 0.3$ G,
$\gamma_{\rm max} = 0.5\times 10^{5}$,
$q_{\rm e} = 2.8 \times 10^{-5}$ cm$^{-3}$ s$^{-1}$,
and $s = 1.4$.
The thick solid curve in Figure \ref{fig:pks2155}
 shows the predicted spectrum
of PKS 2155--304 by numerically solving equations
(\ref{ekin}) and (\ref{pkin}) for
the best parameter set above.
The corresponding electron energy spectrum is shown by the thick solid
curve in Figure \ref{fig:pks2155tot}.
The ratio of $u_{\rm e}/u_{\rm B}$ turns out to be
$u_{\rm e}/u_{\rm B} = 3$ for this model parameters.

As is for Mrk 421 and Mrk 501, the model predictions
for injection rates 50 times
higher and lower than the best fit value
are depicted by dashed and thin solid curves in
Figures \ref{fig:pks2155} and \ref{fig:pks2155tot}.
The numerical values are tabulated in Table \ref{fig:ueub},
and the trend of uncertainties is the same as for Mrk 421 and Mrk 501.

\section{SUMMARY AND DISCUSSION}\label{dis}

To summarize, within the framework of the one-zone synchrotron
self-Compton model, we determined the numerical values of the physical
quantities of TeV blazars Mrk 421, Mrk 501, and PKS 2155--304
in quiescent states.
Those values are searched for by solving the
kinetic equations of electrons and photons taking proper account
of injection, escape, and cooling of electrons and by comparing predicted
radiation spectra with observations.
The best fitted parameters and uncertainties are
estimated.  It is shown that the ratio of the energy density of
electrons to that of magnetic fields can be determined within a
factor of a few and the ratio is about 5, 22, and 3
for Mrk 421, Mrk 501, and PKS 2155--304, respectively.
Thus, the emission region of TeV blazars is particle dominated.
For other parameters, $\delta$ and $\gamma_{\rm max}$ are also determined
within a factor of a few, while the magnetic field strength and the size
have an uncertainty of a factor of 10 and 30, respectively.
Since these results have important implications for
the fundamental understanding of the formation and bulk acceleration
of relativistic jets, below we discuss some of the further
issues to be explored.

\subsection{$u_{\rm e}/u_{\rm B}$ Ratio}

First, we discuss the value of $\gamma_{\rm min}$
because it is important for determining the  
$u_{\rm e}/u_{\rm B}$ ratio. 
From the theoretical standpoint,
in the case of pair plasma jets,
we regard it most likely that $\gamma_{\rm min} \sim \Gamma$,
since the shock first thermalizes a bulk population of particles and
then accelerates them from this pool. 
On the other hand, 
for the normal electron-proton plasma,
there is a wide range of possibilities about $\gamma_{\rm min}$.
One extreme case is that protons and electrons
are separately thermalized, which leads to 
$\gamma_{\rm min}\sim \Gamma$.
Such a separate thermalization is supposed to be 
realized for non-relativistic shocks in supernova remnants.
The other extreme case is that electrons and protons attain
an equilibrium state, which means $\gamma_{\rm min}\sim\frac{m_{\rm
p}}{m_{\rm e}}\Gamma$. This picture is conventionally
assumed for models of cosmic $\gamma$-ray bursts 
(Sari, Piran, \& Narayan 1998).
In this case $\gamma_{\rm min}\sim 10^{4}$ may be realized,
although the real value is likely to be between these two values.

From the observational standpoint, 
it is also difficult to determine $\gamma_{\rm min}$.
Since the observed synchrotron frequency from
an electron with the Lorentz factor $\gamma_{\rm min}$
is given by
\begin{eqnarray}\label{syn,o,min}
\nu_{\rm syn,o,min} =1.2\times 10^{6}B\gamma_{\rm min}^{2}
\frac{\delta}{1+z} \, ,
\end{eqnarray}
electrons with $\gamma_{\rm min} \sim 10$ and $10^{4}$ emit 
synchrotron
photons of $\sim 10^{8}$ and $10^{14}$ Hz, respectively, 
for typical values $B \simeq 0.1$G
and $\delta \simeq 10$.
The former is well below the self-absorption frequency
and the emission in the low frequency band is dominated 
by that from more extended regions.
Since the latter frequency is well above the  self-absorption,
we must modify the model such that emission below $10^{14}$ Hz
is not from X-ray emitting one-zone region 
but from a separate region. 

In our one-zone model, 
to attain the conventional equi-partition state of
$u_{\rm e}=u_{\rm B}$ by simply
changing $\gamma_{\rm min}$ alone,
we need to adopt 
$\gamma_{\rm min} \sim 1.7\times 10^{4}$, 
$3.2\times 10^{4}$, and
$0.5\times 10^{4}$ 
for Mrk 421, Mrk 501, and PKS 2155--304, respectively.
It is of some interest that these analytically estimated
value is near to $10^{4}$
mentioned above.
For $\gamma_{\rm min}\sim 10^{4}$, 
the predicted spectra around near infrared and optical bands
are difficult to match the observed data in the one-zone model.
In this case, whole synchrotron spectra may be
reproduced by a superposition of emission
from inhomogeneous jets.
An example of this kind of modeling is shown in the 
recent work of \cite{ksk01} about the broad-band spectra of Mrk 501
in the flaring stage by a one-zone SSC blob in a conical jet. 
The blob explains X-ray and $\gamma$-ray bands,
while the jet explains the spectrum 
from optical to radio band.
We note that 
their $u_{\rm e}/u_{\rm B}$ ratio for the blob is similar
to ours, irrespective of the actual value of $\gamma_{\rm min}$,
for this flaring stage of Mrk 501 
although they did not explicitly mentioned.

\subsubsection{The Case of $\gamma_{\rm min}=10^{4}$ }

As discussed above,
it is important to examine 
if the alternative case of 
$\gamma_{\rm min}\sim 10^{4}$ and 
$u_{\rm e}/u_{\rm B} \lesssim 1$ 
for normal plasma can reproduce the high energy part of 
the emission spectra in quiescent states. 

\begin{figure}   
  \plotone{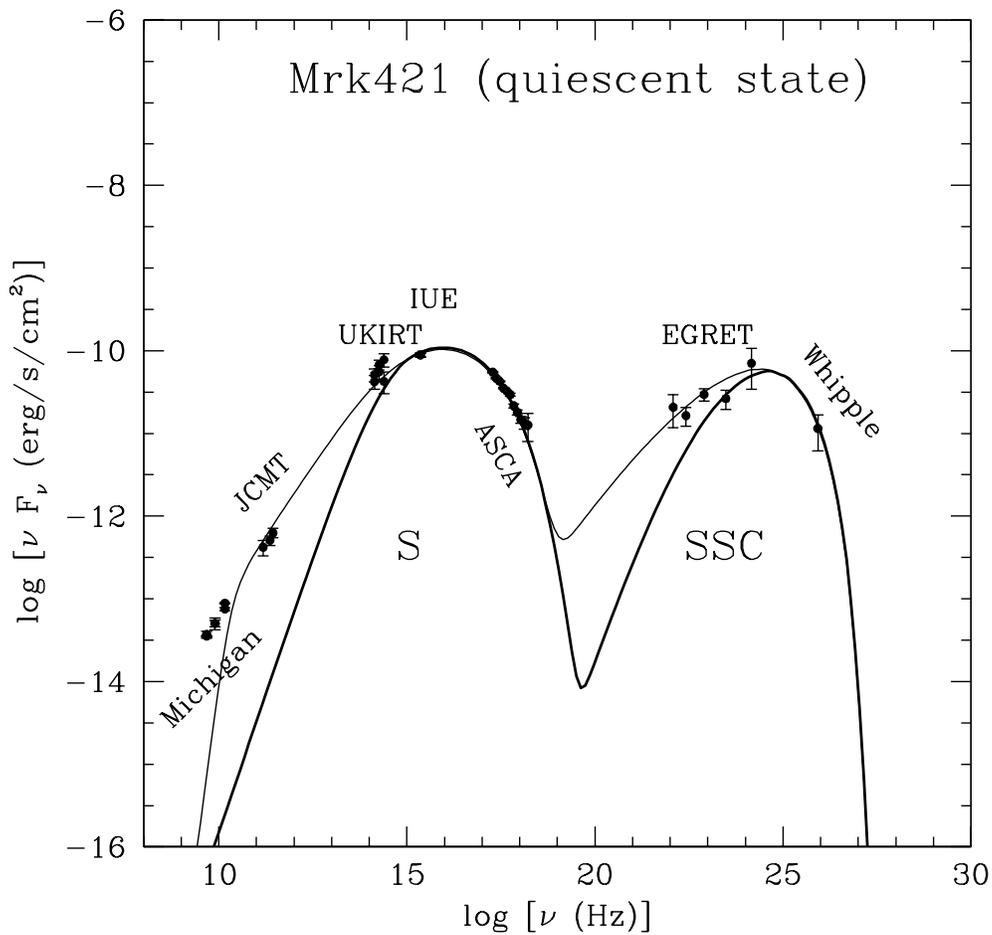}
\caption
{One-zone SSC model spectra for the steady state emission 
of Mrk 421 for $\gamma_{\rm min} = 1 \times 10^{4}$,
and $u_{\rm e}/u_{\rm B} = 2$.
The thick solid line shows the spectrum for this case; 
other parameters are the same as Figure \ref{fig:mrk421}.
The thin solid line shows the best fit spectrum
of Mrk 421 shown in Figure \ref{fig:mrk421}.}
\label{fig:mrk421new}
\end{figure}

\begin{figure}   
  \plotone{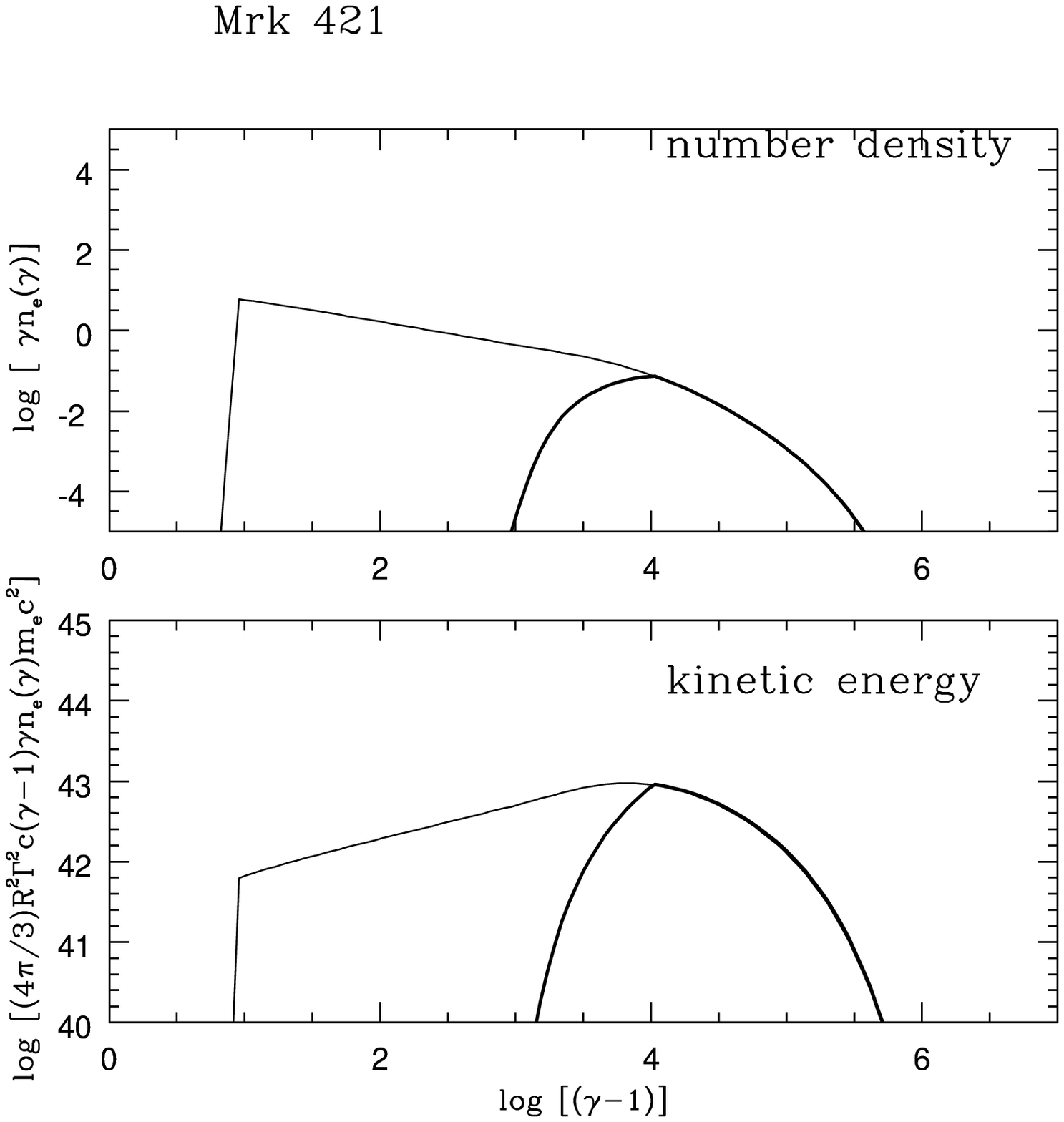}
\caption
{Electron energy spectrum and kinetic power
of Mrk 421 corresponding to Figure \ref{fig:mrk421new}.}
\label{fig:mrk421totnew}
\end{figure}

First, we check the case where 
$\gamma_{\rm br}$ is still
larger than $\gamma_{\rm min}$.
The other parameters are fixed 
as the best fitting ones obtained in the previous section. 
The numerical results for Mrk 421 are shown 
in Figures \ref{fig:mrk421new} 
and \ref{fig:mrk421totnew}.
As is seen in Fig. \ref{fig:mrk421new}, the low energy 
part of both synchrotron and inverse Compton emission is 
short of the observation.
In this case, the numerical value of $u_{\rm e}/u_{\rm B}$
turns out to be 2 and 
yet the emission region is kinetic power dominated. 
Hence we can rule out the possibility of $u_{\rm e}/u_{\rm B}< 1$
for $\gamma_{\rm min}< 10^{4}$.

The second case is $\gamma_{\rm min} > \gamma_{\rm br}$
for which we analytically examine the expected trend.
When $\gamma_{\rm br}< \gamma_{\rm min}$ is satisfied,
the electron energy spectrum is given by
\begin{eqnarray}
n_{\rm e}(\gamma)&=&
q_{\rm e}t_{\rm e,esc} \gamma_{\rm min}^{-s+1}\gamma_{\rm br}
\gamma^{-2}
\, \qquad \qquad {\rm for}\quad
 \gamma_{\rm br}\leq \gamma<\gamma_{\rm min}\nonumber \\
n_{\rm e}(\gamma)&=&
q_{\rm e}t_{\rm e,esc}\gamma_{\rm br}
\gamma^{-s-1}
\qquad {\rm for} \quad \gamma_{\rm min}<\gamma\leq \gamma_{\rm max}.
\end{eqnarray}
This regime is called {\it fast cooling} by \cite{spn98}
in the models of gamma ray bursts. 
In this case, instead of Eq. (\ref{ratio}) we obtain
\begin{eqnarray}
\frac{u_{\rm e}}{u_{\rm B}}
=9\ c_{2}^{-1} c_{3}^{-1}\left(1+\frac{f_{\rm ssc}}{f_{\rm syn}}\right)
\frac{f_{\rm ssc}}{f_{\rm syn}}
\frac{\gamma_{\rm br}}{\gamma_{\rm min}^{s-1}}
\frac
{ \ln (\gamma_{\rm min}/\gamma_{\rm br}) }
{\int^{\gamma_{\rm max}}_{\gamma_{\rm min}}\gamma^{-s+1}d\gamma}  \, .
\end{eqnarray}
Since the combinations of the observables
$\nu_{\rm syn,o,max}$, 
$\nu_{\rm ssc,o,max}$,
$L_{\rm syn,o}$, and
$L_{\rm ssc,o}$ are the same as before, 
from Eqs. (\ref{KN}), (\ref{syn,o,max}), 
and 
(\ref{lratio}),
$\delta\gamma_{\rm max}$, 
$B\delta\gamma_{\rm max}^{2}$, and
$\delta^{4}R^{2}B^{2}$ are not changed.
From Eq. (\ref{tcool}), $\gamma_{\rm br}$ is 
proportional to $R^{-1}B^{-2}$ if we neglect
Compton cooling. 
Thus, we obtain $B\propto\delta\propto\gamma_{\rm br}$,
$R\propto\gamma_{\rm br}^{-3}$, and
$\gamma_{\rm max}\propto\gamma_{\rm br}^{-1}$.
From this, we see that
stronger cooling, i.e., 
smaller values of $\gamma_{\rm br}$,
implies smaller magnetic field
strength and beaming factor and larger size 
and $\gamma_{\rm max}$. 
This is not a favorable choice of parameters, 
because their direction of changes is 
contradictory to the observational facts about strong
beaming and rapid time variability.
Therefore fast cooling regime is not favored.
We thus conclude that our conclusion on kinetic power doninance is 
fairly robust.

\subsection{Proton Components}
It is interesting to know  what constraints on
the jet material can be obtained from this analysis.
Let us assume that the jet consists of relativistic
electrons and cold protons
and that there are neither thermal electrons nor relativistic protons.
Then, using the electron number density for $\gamma_{\rm min} = 10$ and
charge neutrality,
the kinetic power of cold protons
is estimated to be $2.4 \times 10^{44}$ erg s$^{-1}$,
$7.8 \times 10^{44}$ erg s$^{-1}$, and
$4.0 \times 10^{44}$ erg s$^{-1}$, for Mrk 421, Mrk 501, and
PKS 2155--304, respectively.
These values are several to 10 times
larger than the kinetic power of electrons.
This small ratio of the kinetic power of protons to that of electrons,
not withstanding
the large mass ratio,
is due to a large average Lorentz factor of relativistic electrons.
Because the proton kinetic power does not
exceed the Eddington luminosity
for a representative black hole mass in AGN,
we do not make strong argument about jet material in this way alone.
Also, in the case for $\gamma_{\rm min}=10^{4}$, proton kinetic power
is less than that of relativistic electrons. 
Estimation of the large scale kinetic power of
these sources can in principle discriminate between these two possibilities,
i.e., proton-electron jets or electron-positron jets.
Considering the weakness of the extended radio emission of TeV blazars,
the same kind of analysis for GeV blazars seems to be more promising.
We will analyze GeV blazars in the future.

\subsection{Further Comments}
We discuss the injection index $s$.
For TeV blazars studied in this work, $s$ is smaller
than 2.  
Some recent work have reported that
in Fermi acceleration 
at ultra-relativistic shocks $s$ is larger than 2
(\cite{gak99}; \cite{kirk00}), different from our adopted values based on
observed spectra.
This is an interesting open question 
and future research is awaited.  

As for the ratio of the energy densities of the relativistic electrons
to the magnetic fields,
what is physically more meaningful may be
the ratio $(u_{\rm e}+u_{\rm rad}/C_3)/u_{\rm B}$,
because this corresponds to the ratio of the injected kinetic power
to the Poynting power,
while $u_{\rm e}/u_{\rm B}$ corresponds to the remaining ratio
after radiative cooling.  The former ratio becomes 10, 41, and 6
for Mrk 421, Mrk 501, and PKS 2155--304, respectively.
Thus, the conclusion does not change much,
although somewhat strengthened.

As most of the previous work,
we also neglected the correction for
the absorption of TeV $\gamma$-rays due to CIB.
For Mrk 421 and Mrk 501 in quiescent states, this neglect
seems safe, if the observed spectra do not much extend
over the TeV range.  However, for PKS 2155--304 of the redshift of 0.117,
the optical depth for 0.35 TeV photons amounts to about 0.5,
although the exact value depends on the CIB intensity and spectrum.
Thus the intrinsic SSC luminosity of PKS2155--304 should
be a little higher than the estimate given in this paper.
Consequently, the ratio $u_{\rm e}/u_{\rm B}$ is
somewhat higher than the value obtained above and
our conclusion of the kinetic power dominance is
strengthened.

Now, we shortly compare our results with previous work of others.
\cite{tmg98}
derived values of physical parameters
in a similar way to our analytical method.
However, in their paper, electron energy distribution
is not calculated self-consistently.
It means that
normalization, index, and $\gamma_{\rm br}$ of
relativistic electrons are not derived from solving
the cooling and injection processes
but obtained only from a fitting of the observed spectrum
with double power-law spectrum.
Here we examine the case of Mrk 421, because this is the
only source for which multi-frequency spectral fitting was done and
the number density of the accelerated electrons
was estimated by them.

Their best fit parameters for Mrk 421 are
$B = 0.15$ G,
$\delta = 25$,
and $R=2.7\times10^{15}$ cm, and their
electron energy spectrum is given by
$
n_{\rm e}(\gamma)=K\gamma^{-n_{1}}(1+\gamma/\gamma_{\rm br})^{n_{1}-n_{2}}
$, with
$n_{1} = 2.2$, $n_{2} = 4.5$, $\gamma_{\rm br} = 5.6 \times 10^{4}$,
and $K=1.7 \times 10^{5}$.
Note that they are significantly different from our values.
Their electron spectral index 2.2 is steeper than 1.6 in this work,
and it produces a much steeper synchrotron spectrum at low energies.
Their estimate of the emission region is more compact and more strongly
beamed
than ours.
Their parameter values correspond to $u_{\rm e}/u_{\rm B}=371$
for a choice of $\gamma_{\rm min} = 10$ and $\gamma_{\rm max} = 10^{6}$,
and the ratio does not much change for a different choice of
$\gamma_{\rm min}$ and $\gamma_{\rm max}$.
Thus, the
adopted values by \cite{tmg98}
would mean even more kinetic power dominated
states than our results.
Major reason for this difference is due to the choice of $s$,
because replacing $s = 2.2$ by $s = 1.6$ leads to $u_{\rm e}/u_{\rm B} =
19$.

Although we do not make further comparison with other work,
we again emphasize that our method is superior to previous ones
in that the break energy of
electrons is self-consistently determined and
direct spectral fitting is made.

We thank an anonymous referee for useful
comments which help us to improve this paper.
We are grateful to J. Kataoka for kindly providing us with
observational data and for useful discussions.
This work is supported in part by a Grand-in-Aid
from Ministry of Education, Science, Sports and Culture
of Japan (11640236, F. T.).

\begin{deluxetable}{lcrrr}
\tablewidth{33pc}
\tablecaption{Mrk 421 Physical Parameters from SSC Analysis}
\tablehead{
\colhead{}
& \colhead{high injection}
& \colhead{best fit}
& \colhead{low injection} }
\startdata
$\delta$
&  $15.9$
&  $12$
& $ 9.1$
\\
$R$ (cm)
& $ 5.3\times10^{15}$
& $ 2.8\times10^{16}$
& $ 1.4\times10^{17}$
\\
$B$ (G)
&  $ 0.44$
&  $0.12$
&  $0.036$
\\
$\gamma_{\rm max}$
&  $8.3\times10^{4}$
& $1.5\times10^{5}$
& $3.45\times10^{5}$
\\
$s $
& $1.6$
&$1.6$
& $1.6$
\\
$q_{\rm e}$ (cm$^{-3}$ s$^{-1}$)\tablenotemark{a}
& $4.9\times10^{-4}$
& $9.6 \times10^{-6}$
& $1.9 \times10^{-7}$
\\
\hline
$n_{\rm e}^{\rm tot}$ (cm$^{-3}$ )\tablenotemark{a,} \ \tablenotemark{b}
& $1.1\times10^{2}$
& $1.1 \times10^{1}$
& $1.1$
\\
$ \langle\gamma\rangle $\tablenotemark{b}
& $2.3\times10^{2}$
& $3.1\times10^{2}$
& $4.3\times10^{2}$
\\
$\tau_{\rm \gamma \gamma}$\tablenotemark{c}
& $8.9\times10^{-3}$
& $6.9\times10^{-3}$
& $5.5\times10^{-3}$
\\
\hline
$L_{\rm syn,o}$ (erg s$^{-1}$)
& $ 1.3 \times10^{45}$
&  $ 1.3\times10^{45}$
&  $ 1.4\times10^{45}$
\\
$L_{\rm ssc,o}$ (erg s$^{-1}$)
& $0.5\times10^{45}$
&  $0.7\times10^{45}$
&  $0.9\times10^{45}$
\\
$L_{\rm syn+ssc,o}$ (erg s$^{-1}$)
&  $1.8\times10^{45}$
&  $2.0\times10^{45}$
&  $2.3\times10^{45}$
\\
\hline
$L_{\rm poy}$ (erg s$^{-1}$)
& $6.9\times10^{42}$
&  $8.1\times10^{42}$
&  $1.0\times10^{43}$
\\
$L_{\rm e,kin} $ (erg s$^{-1}$)
& $1.8\times10^{43}$
& $4.0\times10^{43}$
& $7.8\times10^{43}$
\\
$L_{\rm syn+ssc}$ (erg s$^{-1}$)
& $ 7.2\times10^{42}$
& $ 1.4 \times10^{43}$
& $ 2.8\times10^{43}$
\\
$L_{\rm p,kin} $ (erg s$^{-1}$)\tablenotemark{d}
& $1.5\times10^{44}$
& $2.4\times10^{44}$
& $3.3\times10^{44}$
\\
\hline
$L_{\rm e,kin}/L_{\rm poy}=u_{\rm e}/u_{\rm B} $
& $3$
& $5$
& $8$
\\
\tablenotetext{a}{$\gamma_{\rm min}$ is fixed at 10.}
\tablenotetext{b}{$n_{\rm e}^{\rm tot}$
and $\langle\gamma\rangle$ are the total number density
and the average Lorentz factor of relativistic electrons, respectively.}
\tablenotetext{c}{$\nu L_{\nu}\simeq1.3\times10^{44}$ erg s$^{-1}$ at 0.3
keV is
adopted in calculating $\tau_{\gamma\gamma}$.}
\tablenotetext{d}{$L_{\rm p,kin} $ is calculated assuming that
the cold proton number density is the same as that of relativistic
electrons.}
\enddata
\end{deluxetable}

\begin{deluxetable}{lcrrr}
\tablewidth{33pc}
\tablecaption{Mrk 501 Physical Parameters from SSC Analysis}
\tablehead{
\colhead{}
& \colhead{high injection}
& \colhead{best fit}
& \colhead{low injection} }
\startdata
$\delta$
&  $14.5$
&  $11$
& $8.3 $
\\
$R$ (cm)
&  $ 2.0\times10^{15}$
&  $ 1.0\times10^{16}$
&  $ 5.3\times10^{16}$
\\
$B$ (G)
&  $0.69 $
&  $0.20$
&  $0.059$
\\
$\gamma_{\rm max}$
& $9.0\times10^{4}$
& $2.0\times10^{5}$
& $3.8\times10^{5}$
\\
$s $
& $1.8$
& $1.8$
& $1.8$
\\
$q_{\rm e}$ (cm$^{-3}$ s$^{-1}$)\tablenotemark{a}
& $8.7 \times10^{-2}$
& $1.7 \times10^{-3}$
& $3.4 \times10^{-5}$
\\
\hline
$n_{\rm e}^{\rm tot}$ (cm$^{-3}$ )\tablenotemark{a,} \ \tablenotemark{b}
& $3.4\times10^{3}$
& $3.4\times10^{2}$
& $3.6\times10^{1}$
\\
$ \langle\gamma\rangle $\tablenotemark{b}
& $1.0\times10^{2}$
& $1.2\times10^{2}$
& $1.5\times10^{2}$
\\
$\tau_{\rm \gamma \gamma}$\tablenotemark{c}
& $2.2\times10^{-2}$
& $1.8\times10^{-2}$
& $1.2\times10^{-2}$
\\
\hline
$L_{\rm syn,o}$ (erg s$^{-1}$)
& $  8.3\times10^{44}$
&  $ 7.8\times10^{44}$
&  $ 8.2\times10^{44}$
\\
$L_{\rm ssc,o}$ (erg s$^{-1}$)
&  $ 9.9\times10^{44}$
&  $ 1.0\times10^{45}$
&  $ 1.2\times10^{45}$
\\
$L_{\rm syn+ssc,o}$  (erg s$^{-1}$)
& $ 1.8\times10^{45}$
& $ 1.8\times10^{45}$
& $ 2.0\times10^{45}$
\\
\hline
$L_{\rm poy}$  (erg s$^{-1}$)
&  $2.0\times10^{42}$
&  $2.4\times10^{42}$
&  $3.4\times10^{42}$
\\
$L_{\rm e,kin} $  (erg s$^{-1}$)
& $3.0\times10^{43}$
& $5.2\times10^{43}$
& $1.1\times10^{44}$
\\
$L_{\rm syn+ssc}$  (erg s$^{-1}$)
&  $ 8.6\times10^{42}$
&  $ 1.5\times10^{43}$
&  $ 2.9\times10^{43}$
\\
$L_{\rm p,kin}  $  (erg s$^{-1}$)\tablenotemark{d}
& $5.2\times10^{44}$
& $7.8\times10^{44}$
& $1.3\times10^{45}$
\\
\hline
$L_{\rm e,kin}/L_{\rm poy}=u_{\rm e}/u_{\rm B} $
& $15$
& $22$
& $32$
 \\
\tablenotetext{a}{$\gamma_{\rm min}$ is fixed at 10.}
\tablenotetext{b}{$n_{\rm e}^{\rm tot}$
and $\langle\gamma\rangle$ are the total number density
and the average Lorentz factor of relativistic electrons, respectively.}
\tablenotetext{c}{$\nu L_{\nu}\simeq7.9\times10^{43}$ erg s$^{-1}$ at 0.3
keV is
adopted in calculating $\tau_{\gamma\gamma}$.}
\tablenotetext{d}{$L_{\rm p,kin}$ is calculated assuming that
the cold proton number density is the same as that of relativistic
electrons.}
\enddata
\end{deluxetable}

\begin{deluxetable}{lcrrr}
\tablewidth{33pc}
\tablecaption{PKS 2155--304 Physical Parameters from SSC Analysis}
\tablehead{
\colhead{}
& \colhead{high injection}
& \colhead{best fit}
& \colhead{low injection} }
\startdata
$\delta$
&  $43.6 $
&  $33$
&  $25.0$
\\
$R$ (cm)
& $ 1.9\times10^{15}$
&  $ 9.0\times10^{15}$
&  $ 4.5\times10^{16}$
\\
$B$ (G)
& $  1.1$
& $0.30$
& $ 0.089$
\\
$\gamma_{\rm max}$
& $2.3\times10^{4}$
& $5.0\times10^{4}$
&  $9.8\times10^{4}$
\\
$s $
& $1.4$
&$1.4$
& $1.4$
\\
$q_{\rm e}$  (cm$^{-3}$ s$^{-1}$)\tablenotemark{a}
& $1.4\times10^{-3}$
& $2.8\times10^{-5}$
& $5.6\times10^{-7}$
\\
\hline
$n_{\rm e}^{\rm tot}$ (cm$^{-3}$ )\tablenotemark{a,} \ \tablenotemark{b}
& $2.5 \times10^{2}$
& $2.4 \times10^{1}$
& $2.3$
\\
$ \langle\gamma\rangle $\tablenotemark{b}
& $2.9\times10^{2}$
& $4.8\times10^{2}$
& $7.3\times10^{2}$
\\
$\tau_{\rm \gamma \gamma}$\tablenotemark{c}
& $5.4\times10^{-3}$
& $4.6\times10^{-3}$
& $3.7\times10^{-3}$
\\
\hline
$L_{\rm syn,o}$  (erg s$^{-1}$)
&  $ 3.4\times10^{46}$
&  $ 3.2\times10^{46}$
& $ 3.4\times10^{46}$
\\
$L_{\rm ssc,o}$  (erg s$^{-1}$)
&  $ 1.1\times10^{46}$
&  $ 1.5\times10^{46}$
& $ 2.0\times10^{46}$
\\
$L_{\rm syn+ssc,o}$  (erg s$^{-1}$)
&  $ 4.5\times10^{46}$
&  $4.7 \times10^{46}$
& $ 5.4\times10^{46}$
\\
\hline
$L_{\rm poy}$  (erg s$^{-1}$)
& $4.1\times10^{43}$
& $4.0\times10^{43}$
& $4.9\times10^{43}$
\\
$L_{\rm e,kin} $  (erg s$^{-1}$)
& $5.2\times10^{43}$
& $1.0\times10^{44}$
& $2.2\times10^{44}$
\\
$L_{\rm syn+ssc}$  (erg s$^{-1}$)
& $ 2.4\times10^{43}$
& $ 4.3 \times10^{43}$
& $ 8.7\times10^{43}$
\\
$L_{\rm p,kin}  $  (erg s$^{-1}$)\tablenotemark{d}
&  $3.3\times10^{44}$
&  $4.0\times10^{44}$
&  $5.7\times10^{44}$
\\
\hline
$L_{\rm e,kin}/L_{\rm poy}=u_{\rm e}/u_{\rm B} $
& $1$
&$3$
& $4$
 \\
\tablenotetext{a}{$\gamma_{\rm min}$ is fixed at 10.}
\tablenotetext{b}{$n_{\rm e}^{\rm tot}$
and $\langle\gamma\rangle$ are the total number density
and the average Lorentz factor of relativistic electrons, respectively.}
\tablenotetext{c}{$\nu L_{\nu}\simeq4.1\times10^{45}$ erg s$^{-1}$ at 0.3
keV is
adopted in calculating $\tau_{\gamma\gamma}$.}
\tablenotetext{d}{$L_{\rm p,kin} $ is calculated assuming that
the cold proton number density is the same as that of relativistic
electrons.}
\enddata
\end{deluxetable}



\end{document}